\documentclass[aps,pra,superscriptaddress,twocolumn,showpacs,nofootinbib,longbibliography]{revtex4-1}
\pdfoutput=1
\usepackage{amssymb,amsmath,amsfonts,amsbsy,amsthm,soul,bm}

\usepackage{dsfont}
\usepackage{bbold}
\usepackage[dvipsnames]{xcolor}

\usepackage{tikz}
\usetikzlibrary{calc}

\usepackage{graphicx}
\usepackage{amsmath}
\usepackage{bbm}
\def\I{ \mathbbm{1} }
\def\Tr{\textrm{Tr}}
\usepackage{hyperref}
\hypersetup{
 colorlinks=true,
 citecolor=[RGB]{0,181,173},
 linkcolor=[RGB]{0,181,173},
  filecolor=[RGB]{239,71,111},      
  urlcolor=[RGB]{0,181,173},}

\def\<{\langle}
\def\>{\rangle}

\newcommand{\var}{\text{Var}}

\begin{document}

\title{Development and Demonstration of an Efficient Readout Error Mitigation Technique for use in NISQ Algorithms}

\author{Andrew Arrasmith}
\email{aarrasmith@rigetti.com}
\affiliation{Rigetti Computing 775 Heinz Ave, Berkeley, California, 94710 USA.\vspace{1em}}

\author{Andrew Patterson}
\email{apatterson@rigetti.com}
\affiliation{Rigetti Computing, 138 Holborn, London, EC1N 2SW, UK.\vspace{0.2em}}

\author{Alice Boughton}
\email{aboughton@rigetti.com}
\affiliation{Rigetti Computing, 138 Holborn, London, EC1N 2SW, UK.\vspace{0.2em}}

\author{Marco Paini}
\email{mpaini@rigetti.com}
\affiliation{Rigetti Computing, 138 Holborn, London, EC1N 2SW, UK.\vspace{0.2em}}

\date{\small\today}
%%%%%%%%%%%%%%%%%%%%%%%%%%%%%%%%%%%%%%%%%%%%%%
%             ABSTRACT
\begin{abstract}
    The approximate state estimation and the closely related classical shadows methods allow for the estimation of complicated observables with relatively few shots. As these methods make use of random measurements that can symmetrise the effect of readout errors, they have been shown to permit simplified approaches to readout error mitigation which require only a number of samples that scales as $\mathcal{O}(1)$ with increasing numbers of qubits. However, these techniques require executing a different circuit at each shot, adding a typically prohibitive amount of latency that prohibits their practical application. In this manuscript we consider the approximate state estimation of readout-mitigated expectation values, and how to best implement that procedure on the Rigetti quantum computing hardware. We discuss the theoretical aspects involved, providing an explicit computation of the effect of readout error on the estimated expectation values and how to mitigate that effect. Leveraging improvements to the Rigetti control systems, we then demonstrate an efficient implementation of this approach. Not only do we find that we can suppress the effect of correlated errors and accurately mitigate the readout errors, we find that we can do so quickly, collecting and processing $10^6$ samples in less than $1.5$ minutes. This development opens the way for practical uses of methods with this type of randomisation.

\end{abstract}

\maketitle

%%%%%%%%%%%%%%%%%%%%%%%%%%%%%%%%%%%%%%%%%%%%%%
%                                                      INTRODUCTION                                                 %
%%%%%%%%%%%%%%%%%%%%%%%%%%%%%%%%%%%%%%%%%%%%%%

\section{Introduction}

The steady progress of improving quantum computing hardware brings us closer and closer to the first practical applications of quantum computing. 
While we fully expect that quantum hardware will advance to the point of allowing the use of fault tolerant quantum algorithms making use of quantum error correction, this may require physical gates with very low error rates and likely at least hundreds of thousands of qubits with high connectivity in order to support practical computations~\cite{kivlichan2020improved,babbush2018encoding}.

Though such quantum hardware has not yet been realized, there is still reason to hope that the noisy, intermediate-scale  quantum (NISQ~\cite{preskill2018quantum}) devices that will be available in the near term may still prove useful for practical applications. Leveraging these NISQ devices will require dealing with hardware error. The earliest practical applications of quantum computers will likely be carefully tailored around the limitations of the hardware and require some form of quantum error mitigation. The field of quantum error mitigation seeks to reduce the impact of hardware errors without having the quantum hardware resources to eliminate them via error correction~\cite{temme2017error,endo2018practical,czarnik2021error,koczor2021exponential,huggins2021virtual,bultrini2021unifying,ferracin2022efficiently}. In this manuscript we focus more narrowly on the problem of mitigating readout errors~\cite{sun2018efficient,chen2019detector,maciejewski2020mitigation,nachman2020unfolding,geller2021conditionally,hamilton2020scalable,geller2021toward,bravyi2021mitigating,seo2021mitigation,karalekas2020quantum,smith2021qubit,van2022model}.

Readout noise encompasses the errors that accrue during the process of performing the quantum measurements and the classification of the results into projected outcomes. Modeling all readout errors as incorrectly categorizing the results, the traditional approach is to construct a transition matrix ($T$) that describes the mapping between the ideal probability distribution and the observed one that has been impacted by the state preparation and measurement errors~\cite{sun2018efficient}. If state preparation errors are rigorously removed or negligible compared to measurement errors, an estimate of the ideal readout probabilities can be computed by multiplying the measured distribution with an appropriately constrained estimate of the inverse of this $T$ matrix ~\cite{geller2021conditionally,sun2018efficient,chen2019detector,maciejewski2020mitigation}.  There are also alternative approaches, such as Bayesian unfolding, that allow one to avoid the explicit need for computing this pseudo-inverse~\cite{nachman2020unfolding}. This classical miscategorization error mitigation model is fairly effective as these classical errors appear to be the dominant source of readout error, though it does not fully account for the possibility of coherent errors during the readout process~\cite{maciejewski2020mitigation}.

The primary difficulty with these approaches is that the amount of data required to implement these strategies quickly becomes prohibitive with larger systems as they involve the estimation of operators acting on exponentially large probability spaces. A number of approaches have therefore been proposed in order reduce the requirement for data. One such approach is to assume that the readout errors are not correlated (or have limited correlations) between qubits and estimate a collection of operators acting on small sets of qubits ~\cite{hamilton2020scalable,geller2021toward,bravyi2021mitigating}. While this assumption may fail to be satisfactory in cases where high accuracy is needed, it may provide a practical benefit as it changes the resource scaling from exponential to linear in the number of qubits being measured.

Aside from approximations that assume the absence of correlations, the amount of data required can also be reduced by imposing symmetries on the readout error channel. One approach that has been used is to randomly choose whether or not to apply Pauli $X$ gates (bit-flips) on each qubit immediately before measurement and then flip the corresponding bits in the measurement results~\cite{karalekas2020quantum,smith2021qubit}. This symmetrization results in the $T$ matrix being symmetric about its diagonal and anti-diagonal, giving a substantial reduction in the number of elements that need to be measured but not eliminating the exponential scaling. Further symmetrization is possible if instead of only having a chance of applying $X$ gates we also randomly apply other single-qubit Pauli gates as well, as done in~\cite{beale2020true}, transforming coherent errors from readout into stochastic Pauli errors and removing the problem of coherent readout errors.

When working with bit-flip symmetrized readout errors, it has been shown that the readout error can be seen as suppressing expectation values by some factor~\cite{karalekas2020quantum,smith2021qubit,van2022model}. This approach takes the number of calibrations needed to approximately invert the noise from scaling exponentially with the number of qubits to scaling linearly with the number of independent measurement bases needed for the computation. This difference makes readout error mitigation practical for computational problems of any size, provided that they only involve tractable (i.e. scaling polynomially with the number of qubits) numbers of measurement bases. 

In this paper, we consider a higher degree of symmetrization which is achieved using the approximate state tomography formalism~\cite{paini2000quantum,paini2019approximate,paini2021estimating}. The approximate state formalism, largely equivalent to the classical shadows formalism~\cite{huang2020predicting}, allows one to estimate expectation values from a collection of randomly selected measurements. In the context of the classical shadow formalism it has been shown that it is possible to perform the calibration for any number of measurement bases using only a single state preparation with randomised measurements~\cite{koh2020classical,chen2021robust}. So long as one works with sufficiently local observables, the approximate state or classical shadows formalism scales efficiently~\cite{paini2019approximate,huang2020predicting}. Working with this framework we study the effects of measurement errors on readout values, finding that, unlike previous claims~\cite{karalekas2020quantum,chen2021robust}, the randomisation of readout is not actually equivalent to a twirl~\cite{bennett1996mixed,dankert2009exact} though it is similar. Additionally, we show how using the higher degree of symmetry in the sampling suppresses the impact of asymmetric correlated readout errors, making a tensor product approximation exceedingly good at mitigating errors on current devices. Finally, we demonstrate a fast implementation of readout error mitigation with the approximate state formalism on the Rigetti hardware stack. 

The remainder of the paper is structured as follows. We begin by reviewing the approximate state formalism for estimating expectation values. Next, we discuss the effect of the approximate state method on readout error channels and our error mitigation approach. We then discuss the efficient method by which we implement this technique on the Rigetti control systems and quantum processor's. Finally, we show the results of applying our readout error mitigation on Rigetti's superconducting quantum processors. Specifically, we examine the method's speed and ability to break error-induced correlations in readout. We then consider an example problem where we use this method to mitigate the output simulations of plasma physics~\cite{shi2021simulating}. We conclude by discussing possible applications and extensions of the method in the context of near-term quantum computation.
%including estimating energies in VQE~\cite{peruzzo2014variational}

%%%%%%%%%%%%%%%%%%%%%%%%%%%%%%%%%%%%%%%%%%%%%%
%                                                 BACKGROUND                                                         %
%%%%%%%%%%%%%%%%%%%%%%%%%%%%%%%%%%%%%%%%%%%%%%

\section{The Approximate State Formalism}  \label{Approximate State/Classical Shadows Formalism}
\subsection{Approximate State Tomography}\label{sec:approximate_state}
The central idea of the approximate state formalism is to expand a density matrix using kernel operators whose coefficients are probabilities. These kernel operators $K(m,x)$, where $x$ determines the measurement taken and $m$ is a possible measurement outcome, must constitute a tomographically complete set. 

We follow the original approximate state formalism paper in focusing on kernels based on the full group of single-qubit unitaries, $SU(2)$, of the measurement basis~\cite{paini2019approximate}. We note that this formalism, with different choices of groups and their representations, can be generalized to working with kernel operators based on measurements of more general sets of positive, operator value measures (POVMs) so long as the set of kernel operators remains tomographically complete~\cite{paini2000quantum}

Choosing a standard representation of $SU(2)$ and a group given by the direct product of $SU(2)$ $Q$ times (corresponding to $Q$ qubits), the kernel operators are tensor products of single-qubit kernel operators:
\begin{equation}
    K(\{m,\vec{n}\}) = \bigotimes_{j=1}^Q K_1(m_j,\vec{n}_j)
\end{equation}
with
\begin{equation}
    K_1(m_j,\vec{n}_j) = \frac{1}{2}\left(\I + 3 m_j \vec{\sigma}\cdot \vec{n}_j\right).
\end{equation}
Here the set of unit vectors $\{\vec{n}\}$ specify the Bloch sphere direction of the measurement axis for each qubit and the measurement outcomes $\{m\vert m\in\{-1,1\}\}$ are eigenvalues of the single-qubit measurement operators. Here and below we drop the explicit indices and write $\{m,\vec{n}\}$ to denote the set of $Q$ pairs of single-qubit outcomes and measurement directions.

A density matrix $\rho$ can be decomposed onto these kernel operators as
\begin{align}
    \rho =& \prod_{j=1}^Q\left(\int_{\Sigma} d\mu(\vec{n}_j) \sum_{m_j\in \{-1,1\}}\right)  \nonumber \\
    &\qquad \Tr\left[ \vert \{\vec{n}, m\} \rangle \langle \{\vec{n}, m\} \vert\rho\right]K(\{m,\vec{n}\})\nonumber\\
    =& \prod_{j=1}^Q\left(\int_{\Sigma} d\mu(\vec{n}_j) \sum_{m_j\in \{-1,1\}}\right) p(\{m,\vec{n}\}) K(\{m,\vec{n}\}).
\end{align}
Here the integration measure is uniform over the directions to be measured on each qubit's Bloch sphere. The single-qubit projectors $\vert \vec{n}_j, m_j \rangle \langle \vec{n}_j,m_j \vert$ are onto the eigenbasis of $\sigma\cdot \vec{n}_j$. That is, the measurement basis is rotated from the Pauli $Z$ eigenbasis by a single-qubit unitary $U_1(\vec{n}_j)$, which is chosen so that the state $U_1(\vec{n}_j)|0\rangle$ points along the Bloch sphere direction specified by $\vec{n}_j$.

While reconstructing a full density matrix does not scale efficiently, this expression allows us to compute the expectation value of an operator $O$ as
\begin{align}\label{eq:estimator}
    \Tr[\rho O] = \prod_{j=1}^Q&\left(\int_{\Sigma} d\mu(\vec{n}_j) \sum_{m_j\in \{-1,1\}}\right)\nonumber \\
    &p(\{m,\vec{n}\}) \Tr[K(\{m,\vec{n}\})O].
\end{align}
Practical use of this formalism is then facilitated by Monte Carlo estimation of such expectation values. Denoting the estimator for $\Tr[\rho O]$ as $\overline{R[O]}$, we construct that estimator as~
\cite{paini2019approximate,paini2021estimating}
\begin{align}\label{eq:montecarlo_estimator}
    \overline{R[O]} =& \frac{1}{N}\sum_{a=1}^N \Tr\left[ K(\{m,\vec{n}\}_a) O\right]\nonumber\\
    =& \frac{1}{N}\sum_{a=1}^N R[O]_a.
\end{align}
Here the overline on $\overline{R[O]}$ denotes the sample mean of the single-measurement estimators $R[O]$ for $\Tr[\rho O]$ and $N$ is the total number of repeated measurements included in that sample. For each measurement repetition we are uniformly drawing the directions $\{\vec{n}\}$, preparing and measuring the quantum state $\rho$ in the associated bases, and recording the outcomes $\{m\}$.

Given the form of the kernel operators being used, it is convenient to work in terms of the decomposition of the operator $O$ into a sum over tensor products of Pauli operators (Pauli strings):

\begin{equation}
    O = \sum_{\vec{i}}c_{\vec{i}}P_{\vec{i}}.
\end{equation}
Here $c_{\vec{i}}$ is the (real) coefficient multiplying the Pauli string $P_{\vec{i}}$ which is defined as
\begin{equation}
    P_{\vec{i}} = \bigotimes_{j=1}^Q \sigma_{i_j}.
\end{equation}
We adopt the convention that $\sigma_0$, $\sigma_1$, $\sigma_2$, and $\sigma_3$ denote the single-qubit identity, Pauli $X$, Pauli $Y$ and Pauli $Z$ operators, respectively.

The expectation value estimator for an operator $O$ with $N$ samples has a variance bounded by~\cite{paini2021estimating}:
\begin{equation}\label{eq:var_bound}
    \var\left( \overline{R[O]}\right) \leq \frac{\|O\|^2}{N}  
\end{equation}
where $\|\cdot\|$ is a seminorm. For single-qubit operators, this seminorm is
\begin{equation}
    \|O\|=\sqrt{3\sum_{i=1}^3 c_i^2},
\end{equation}
and for multi-qubit operators this expression becomes:
\begin{equation}\label{eq:spherical_seminorm}
    \|O\|=\sqrt{\sum_{\vec{i},\vec{k}\neq \vec{0}} 3^{r_{\vec{i},\vec{k}}} \Delta_{\vec{i},\vec{k}} |c_{\vec{i}}||c_{\vec{k}}|}.
\end{equation}
Here $r_{\vec{i},\vec{k}}$ is the number of qubit indices $\ell$ such that $i_\ell \neq 0$ and $k_\ell\neq 0$ for the pair $\vec{i},\vec{k}$. $\Delta_{\vec{i},\vec{k}}=0$ if there exists some qubit index $\ell$ such that $i_\ell\neq 0$, $k_\ell \neq 0$ and $i_\ell\neq k_\ell$, otherwise $\Delta_{\vec{i},\vec{k}}=1$~\cite{paini2021estimating}.

\subsection{Sampling with t-designs}
Rather than integrating over measurement directions on a Bloch sphere, we can translate this procedure into sampling over single-qubit $t\geq2$ designs, such as the single-qubit Clifford group (also called the octahedral group~\cite{barends2014rolling}) typically used in the classical shadows literature~\cite{huang2020predicting}. To arrive at this conclusion, note that Equation~\eqref{eq:estimator} for measuring an expectation value can be re-expressed as an integral over single-qubit Haar measures. More explicitly,
\begin{align}
    \Tr[\rho O] =& \prod_{j=1}^Q\left(\int_{\Sigma} d\mu(\vec{n_j}) \sum_{m_j\in \{-1,1\}}\right)\nonumber \\
    &\quad \Tr\left[\vert \{\vec{n}, m\} \rangle \langle \{\vec{n}, m\} \vert \rho\right]\nonumber \\
    &\quad\Tr\left[K(\{m,\vec{n}\})O\right]\nonumber\\
    =& \prod_{j=1}^Q\left(\int_{\Sigma} d\mu(\vec{n_j}) \sum_{m_j\in \{-1,1\}}\right)\nonumber \\
    &\quad\Tr\left[U(\{\vec{n}\})^{\dagger}\vert \{\hat{z}, m\} \rangle \langle \{\hat{z}, m\} \vert
     U(\{\vec{n}\})\rho \right]\nonumber \\
    &\quad \Tr\left[U(\{\vec{n}\})^{\dagger}K(\{m,\hat{z}\})U(\{\vec{n}\})O\right].
\end{align}
We note that the mapping from Bloch sphere directions $\vec{n}$ to the associated rotation unitaries $U(\vec{n})$ is under-constrained in the approximate state formalism as both the projectors $U(\{\vec{n}\})^{\dagger}\left(\bigotimes_{j=1}^Q|m_j,\hat{z}\rangle\langle m_j\hat{z}|\right)U(\{\vec{n}\})$ and the the kernel operators $U(\{\vec{n}\})^{\dagger}K(\{m,\hat{z}\})U(\{\vec{n}\})$ are unchanged under the transformation 
\begin{align}
    U(\{\vec{n}\})\rightarrow &U'(\{\vec{n},\psi\})\nonumber\\
    &=\left(\bigotimes_{j=1}^Q \mathcal{R}_z(\psi_j)\right)U(\{\vec{n}\}).
\end{align}
Here $\mathcal{R}_z(\psi)$ is a single-qubit rotation about the $\hat{z}$ axis. With this freedom in mind, given that we work with single-qubit rotations ($U(\{\vec{n}\})=\bigotimes_{j=1}^Q U_1(\vec{n}_j)$) we then have
\begin{align}\label{eq:haar_estimator}
    \Tr\left[\rho O\right] =&  \left(\prod_{j=1}^Q \int_{SU(2)}d\mu'(U_{1,j}')\sum_{m_j\in \{-1,1\}}\right)\nonumber \\
    &\qquad\Tr\left[\left(\bigotimes_{j=1}^Q U_{1,j}'^{\dagger}|m_j,\hat{z}\rangle\langle m_j,\hat{z}|U_{1,j}'\right)\rho \right]\nonumber\\ &\qquad\Tr\left[\left(\bigotimes_{j=1}^Q U_{1,j}'^{\dagger}K(m_j,\hat{z}_j)U_{1,j}'\right)O\right].
\end{align}
Here we have made the change of variables $\vec{n_j},\psi\to U_{1,j}=U_1'(\vec{n_j},\psi)$ and the integration measure $d\mu'(U_{1,j})$ is the Haar measure on $SU(2)$. 

As Equation~\eqref{eq:haar_estimator} is an integral on the Haar measure, by definition of a $t$-design~\cite{dankert2009exact} the integration can be replaced with averaging over $t$-designs so long as $t$ is large enough. Going forward, we will require a $t\geq2$ design, as the above expression has terms with the second power of the unitaries $U'_j$. For discussion on extending the above arguments to the computation of variances see Appendix~\ref{app:t-design}.

Below, we will make use of the tetrahedral group for sampling, as it forms a $2$-design on the space of single-qubit unitaries with only $12$ elements~\cite{barends2014rolling}. We note, however, that applications that involved computing higher than second moments would require sampling from higher order designs.

We note a subtle point. Using a group that is a $t$-design to do the Monte Carlo sampling in Equation~\eqref{eq:montecarlo_estimator} is distinct from deriving the kernel operators $K(m,x)$ associated with that group. In general using a different group will result in kernel operators that look different and correspond to a different set of measurements.

Finally, we remark that while the use of $t$-designs here is useful for setting up the derivation of Section~\ref{sec:The Randomised Readout Channel}, it is not the only approach. As discussed in Appendix~\ref{app:twirl}, one could arrive at similar results from the irreducibility condition of considering irreducible representations of finite subgroups of $SU(2)$.

\subsection{Non-Uniform Sampling from SU(2)}\label{sec:non_uniform}
As described in Section~\ref{Approximate State/Classical Shadows Formalism}, sampling measurement directions uniformly from the Bloch sphere for each qubit provides an expectation value estimator $\overline{R[O]}$ with a variance that depends only on the structure of $O$. Specifically, as a byproduct of phrasing that variance calculation as a set of Haar integrals, we can immediately see that the variance is unchanged by transforming $O$ with tensor products of single-qubit rotations. 

This symmetric variance is often desirable when working with complex operators. For such operators, we need to consider the expectation values of many non-commuting Pauli strings, as all Pauli strings with the same number of identity elements are estimated with equal precision. However, it is possible to bias the sampling in such a way that measurements along some direction $\vec{n}$ have a smaller variance. The cost of biasing the measurements like this is increasing the variance of orthogonal measurements. Here we demonstrate a simple example of this in the case of choosing to reduce the variance of measurements along $\hat{z}$ axis of the Bloch sphere. We note that inserting a fixed tensor product of single-qubit unitaries can reorient the bias of this sampling to any other tensor product basis if desired.

To begin with, we describe the measurement directions $\vec{n}$ in terms of spherical coordinates. We adopt the convention that $\phi$ is the azimuth angle and $\theta$ is the polar angle. A simple way to introduce non-uniform sampling on this sphere is to sample $\phi$ uniformly from the interval $[0,2\pi]$ and sample $\theta$ uniformly from the interval $[0,\pi]$, causing the sampled points to concentrate on the poles of the Bloch sphere. We will therefore refer to this sampling method as pole-concentrated sampling.

One motivation for the consideration of this non-uniform sampling on the Bloch sphere is that directly sampling from these intervals with a pseudo-random number generator is more efficient than uniform sampling when performed in control systems close to the QPU as it does not require the use of the inverse trigonometric functions the spherically symmetric version does. Therefore, this sampling method may be advantageous for cases with simple asymmetric operators and where the execution speed is critical. See Section~\ref{sec:hardware} for more details on how such an implementation can be achieved. Additionally, if the structure of the operator being sampled is such that certain single-qubit operators appear more frequently, such a biased sampling may serve to reduce the variance of the estimate. 

The estimator in Equation~\eqref{eq:estimator} can be modified to accomplish this pole-concentrated sampling as:
\begin{align}\label{eq:non-uniform estimator}
    \Tr[\rho O] =& \prod_{j=1}^Q\left(\int_0^\pi \int_0^{2\pi}\frac{d\theta_j d\phi_j} {2\pi^2}\frac{\pi}{2}\textrm{sin}(\theta_j)  \sum_{m_j\in \{-1,1\}}\right)\nonumber \\
    &\qquad p\left(\{m,\vec{n}(\theta,\phi)\}\right) \Tr[K\left(\{m,\vec{n}(\theta,\phi)\}\right)O]\nonumber\\
    =& \prod_{j=1}^Q\left(\int_0^\pi \int_0^{2\pi}\frac{d\theta_j d\phi_j} {2\pi^2}  \sum_{m_j\in \{-1,1\}}\right)\nonumber \\
    &\qquad p\left(\{m,\vec{n}(\theta,\phi)\}\right) \Tr[K'\left(\{m,\vec{n}(\theta,\phi)\}\right)O].
\end{align}
Here $K_1'(m,\vec{n}(\theta,\phi)) \equiv \frac{\pi}{2}\textrm{sin}(\theta)K_1(m,\vec{n}(\theta,\phi))$ is the single-qubit kernel associated with this uniform interval sampling. This factor of $\frac{\pi}{2}\textrm{sin}(\theta)$ comes from the spherically symmetric integration measure. 

With this modified kernel we can define a Monte Carlo estimator similar to the one in Equation~\eqref{eq:montecarlo_estimator}:
\begin{align}\label{eq:non-uniform_montecarlo_estimator}
    \overline{R'[O]} =& \frac{1}{N}\sum_{a=1}^N \Tr\left[K'(\{m,\vec{n}\}_a) O\right].
\end{align}
This modified estimator essentially re-weights the samples by factors of  $\frac{\pi}{2} \textrm{sin}(\theta)$ for each qubit in order to achieve the same mean while sampling from a different distribution than the estimator in Equation~\eqref{eq:montecarlo_estimator}.

The concentration of samples around the poles from this approach leads to different variances for different Pauli operators. Specifically, this sampling method results in the following bounds on the single-qubit variances (see Appendix~\ref{app:pole_concentrated} for details):
\begin{equation}
    \textrm{Var}\left(\overline{R'[Z]}\right) \leq \frac{9 \pi^2}{32N},
\end{equation}
\begin{equation}
    \textrm{Var}\left(\overline{R'[X]}\right) \leq\frac{27 \pi^2}{64N},
\end{equation}
and
\begin{equation}
    \textrm{Var}\left(\overline{R'[Y]}\right)  \leq\frac{27 \pi^2}{64N}.
\end{equation}
This method also results in the expectation value of the identity having a non-zero variance as each single-qubit, single-measurement estimate depends on the value of $\textnormal{sin}(\theta)$ that was drawn. We therefore have
\begin{equation}
    \textrm{Var}\left(\overline{R'[\I]}\right) = \frac{\pi^2}{8N}.
\end{equation}

More generally, we can introduce a new seminorm that reflects this asymmetry and use that to bound the variance of estimators using this pole-concentrated sampling. For more details, see Appendix~\ref{app:pole_concentrated}.

\subsection{Relationship to Classical Shadows}\label{sec:classical_shadow}
The term classical shadow tomography is more commonly discussed than approximate state tomography in the literature, but they are essentially different descriptions of the same process of using randomised measurements~\cite{paini2019approximate,huang2020predicting}. With classical shadow tomography one samples a unitary $U$ from some ensemble $\mathcal{U}$, such as the single-qubit Clifford group, and constructs an unbiased estimator $\widehat{\rho}$ for the density matrix $\rho$ from the measured bitstring $|b\rangle$:
\begin{equation}
    \widehat{\rho} = \mathcal{M}^{-1}(U^\dagger |b\rangle\langle b| U).
\end{equation}
The reconstruction map $\mathcal{M}^{-1}$ is generally a non-physical channel, but it can be applied in classical memory after the measurement~\cite{huang2020predicting}. In the presence of readout errors, this reconstruction map and its inverse $\mathcal{M}$, called the shadow channel, is constructed using the ensemble $\mathcal{U}$ as well as the readout error channel $\mathcal{E}$. Specifically,~\cite{koh2020classical,chen2021robust}
\begin{equation}
    \mathcal{M}(\cdot) = \mathbbm{E}\left[U^\dagger\left(\sum_b |b\rangle\langle b|\mathcal{E}\left(\mathbbm{E} \right[ U \cdot U^\dagger\left]\right)|b\rangle\langle b|\right) U\right].
\end{equation}

The effect of this error channel $\mathcal{E}$ can only be mitigated if the inverse of this noisy $\mathcal{M}$ is known~\cite{koh2020classical,chen2021robust}. However, it has been asserted that the randomised readout is a twirling operation and thus significantly reduces the number of measurements needed to characterize the effect of $\mathcal{E}$ on $\mathcal{M}$. While we will show below that this assertion is not precisely correct, we find that the conclusions drawn from it about the reduction in degrees of freedom that need to be measured still hold. 

%%%%%%%%%%%%%%%%%%%%%%%%%%%%%%%%%%%%%%%%%%%%%%
%                            				THE RANDOMISED READOUT CHANNEL                                        %
%%%%%%%%%%%%%%%%%%%%%%%%%%%%%%%%%%%%%%%%%%%%%%

%\maketitle
\section{Approximate State Tomography with Noise} \label{sec:The Randomised Readout Channel}
The high degree of symmetrization involved in the process of an approximate state experiment means that the effective readout error channels that result can be far simpler than the generic physical readout error channels they arise from. In fact, we find that, as posited previously~\cite{chen2021robust}, the effect of noise simplifies to the point where it can be approximately inverted with random measurements of a single state. 

To arrive at that result, we begin with Equation~\eqref{eq:haar_estimator} and rearrange to get (dropping the primes on the $U_j$'s and on the measure, understanding the $U_j$'s to be single-qubit unitaries):
\begin{align}
    \Tr[\rho O] =& \left(\prod_{j=1}^Q \int_{SU(2)}d\mu(U_j)\right)\nonumber\\
    &\;Tr\Bigg[\left(\bigotimes_{j=1}^Q|m_j,\hat{z}\rangle\langle m_j,\hat{z}|\right)\nonumber\\
    &\qquad\left(\bigotimes_{j=1}^Q U_j\right)\rho\left(\bigotimes_{j=1}^Q U_j^{\dagger}\right) \Bigg]\nonumber \\ &\;\Tr\left[\left(\bigotimes_{j=1}^Q U_j^{\dagger}K(m_j,\hat{z}_j)U_j'\right)O\right].
\end{align}

As the physical implementation of this formalism involves rotating the state before taking a measurement, an arbitrary quantum channel $\mathcal{E}$ acting on the state at measurement (i.e. after the $U_j$ rotations) distorts the expectation value $\Tr[\rho O]$ into the function $f(\mathcal{E},\rho,O)$:
\begin{align}\label{eq:noise_distortion}
    f(\mathcal{E},\rho,O)=&\left(\prod_{j=1}^Q \int_{SU(2)}d\mu(U_j)\right)\nonumber\\ &\;Tr\Bigg[\left(\bigotimes_{j=1}^Q|m_j,\hat{z}\rangle\langle m_j,\hat{z}|\right)\nonumber\\
    &\qquad\mathcal{E}\left(\left(\bigotimes_{j=1}^Q U_j\right)\rho\left(\bigotimes_{j=1}^Q U_j^{\dagger}\right)\right) \Bigg]\nonumber\\ &\;\Tr\left[\left(\bigotimes_{j=1}^Q U_j^{\dagger}K(m_j,\hat{z}_j)U_j'\right)O\right]
\end{align}

In order to capture the effect of this distortion, we will now focus on the Pauli strings as any $O$ can be decomposed into Pauli strings. Repeatedly utilizing a Haar integration identity~\cite{puchala2017symbolic}, it can be shown that for any Pauli string $P_{\vec{i}}$, the distorted result $f(\mathcal{E},\rho,P_{\vec{i}})$ is related to the true expectation value $\Tr[\rho P_{\vec{i}}]$ as 
\begin{equation}\label{eq:suppression}
    f(\mathcal{E},\rho,P_{\vec{i}}) = \frac{1}{2^Q}\Tr[M_{\vec{i}}\mathcal{E}(M_{\vec{i}})]\Tr[\rho P_{\vec{i}}]
\end{equation}
with
\begin{equation}\label{eq:Mdef}
    M_{\vec{i}}=\bigotimes_{j=1}^Q\left(\delta_{i_{j}0}\I +\left(1-\delta_{i_{j}0}\right)\sigma_z\right).
\end{equation}
See Appendix~\ref{app:effective_error_channel} for the derivation of this result. In other words, the expectation value of a Pauli string is suppressed by a factor that depends only on the properties of the error channel and the location of non-identity Paulis in that string.

\subsection{Relationship to Twirling Operations}

We note that the result in Equation~\eqref{eq:suppression} is very similar to the action of a twirl with the same group. For the case of a twirl the suppression of the expectation value depends on the average of $3^{N_{P_{\vec{i}}}}$ diagonal elements in the Pauli transfer matrix representation of $\mathcal{E}$ rather than just the one associated with $M_{\vec{i}}$. Here $N_{P_{\vec{i}}}$ in the number of non-identity terms in the Pauli string $P_{\vec{i}}$. See Appendix~\ref{app:twirl} for more details.

The significance of this averaging in the twirl operation is that, while the effect of randomised readout is very similar to that of twirling the error channel, \emph{the suppression factors are not equal in general}. However, even though randomised readout does not quite implement a twirling operation, its action only differs by a multiplicative factor. Thus, the incorrect assertions that it is a twirl in previous literature~\cite{karalekas2020quantum,chen2021robust} still led to the construction of valid error mitigation protocols in those works as they directly measure the suppression factors rather than calculating them.

\subsection{Tensor Product Error Models}
Finally, we remark that for the case of an error channel with a tensor product structure ($ \mathcal{E}=\bigotimes_{k=1}^Q e_k$) we have
\begin{align}
    \frac{1}{2^Q}\Tr\left[M_{\vec{i}}  \mathcal{E}\left(M_{\vec{i}}\right)\right] = \prod_{j=1}^Q \Big( \delta_{i_j,0}&+(1-\delta_{i_j,0})\nonumber\\
    &\cdot\frac{1}{2}\Tr[\sigma_z e_j(\sigma_z)]\Big).
\end{align}
Since we end up with a product of single-qubit factors, we can describe this symmetrized effective channel as a tensor product of single-qubit depolarizing channels, as the effect is to suppress all expectation values by a product of single qubit factors. 

We note that as the effective channel only depends on $\frac{1}{2^Q}\Tr\left[M_{\vec{i}}  \mathcal{E}\left(M_{\vec{i}}\right)\right]$ and not the entire channel $\mathcal{E}$, the effective channel does not include any impact from asymmetric correlated errors. Therefore, when there is an expectation that most of the correlated error is asymmetric under single-qubit unitary transformations, it is reasonable to expect that a tensor product approximation to the effective channel may be justified. As we will see in Section~\ref{Experimental Results}, approximating the effective error channel as such a tensor product can be very accurate on Rigetti's current hardware.

%%%%%%%%%%%%%%%%%%%%%%%%%%%%%%%%%%%%%%%%%%%%%%
%                                                        Readout Error Mitigation                                       %
%%%%%%%%%%%%%%%%%%%%%%%%%%%%%%%%%%%%%%%%%%%%%%

%\maketitle
\section{Readout Error Mitigation} \label{Readout Error Mitigation}

The procedure we advocate for mitigating readout errors in the approximate state formalism is essentially the same as the method proposed in~\cite{chen2021robust} in the classical shadows language. Here we list the steps of our approach in the approximate state formalism.

The basic approach to estimating the error mitigated expectation value of an observable  $O = \sum_{\vec{i} }c_{\vec{i}}P_{\vec{i}}$ with a state $\rho$ is as follows. For brevity we only show the spherically symmetric sampling case, the generalization to other cases like the pole-concentrated method is the straightforward replacement of the single shot estimators as appropriate.
\begin{enumerate}\label{list:error_mitigation_steps}
    \item \emph{Noisy Term Estimation:} 
    \begin{enumerate}
        \item Prepare and measure the state $\rho$ a number of times $N_{\rho}$, each time applying randomly drawn unitary rotations on each qubit immediately prior to measurement.
        \item For each non-zero $c_{\vec{i}}$, record the sample mean of the noisy, single shot estimators as
        \begin{align}
            \hat{f}(\mathcal{E},\rho,P_{\vec{i}}) = \frac{1}{N_{\rho}}\sum_{a=1}^{N_{\rho}}& \prod_{j=1}^Q\Big(\delta_{i_j,0}\nonumber\\
            &+(1-\delta_{i_j,0})3 m_{j,a} \vec{n}_{j,a}\cdot\hat{i_j}\Big).
        \end{align}
        Here $\hat{i_j}$ is the unit vector corresponding to a measurement of $\sigma_{i_j}$.
        These $\hat{f}(\mathcal{E},\rho,P_{\vec{i}})$  are estimators for the distorted expectation values $f(\mathcal{E},\rho,P_{\vec{i}})$ (defined in Equation~\eqref{eq:noise_distortion}).
    \end{enumerate}
    \item\label{step:suppression} \emph{Noise Suppression Estimation:}
    \begin{enumerate}
        \item Prepare and measure the all zeros state $\bigotimes_{j=1}^Q|0\rangle\langle0|$ a number of times $N_c$ with randomly drawn unitary rotations immediately prior to the measurement.
        \item For each distinct $M_{\vec{i}}$ associated with at least one non-zero $c_{\vec{i}}$ record the sample mean of the noisy, single shot estimators as
        \begin{align}
            \hat{f}(\mathcal{E},\bigotimes_{j=1}^Q|0\rangle\langle0|,M_{\vec{i}}) = \frac{1}{N_{c}}\sum_{a=1}^{N_c}& \prod_{j=1}^Q\Big(\delta_{i_j,0}\nonumber\\
            &+(1-\delta_{i_j,0})3 m_{j,a} \vec{n}_{j,a}\cdot\hat{z}\Big).
        \end{align} 
        As above, these $\hat{f}(\mathcal{E},\bigotimes_{j=1}^Q|0\rangle\langle0|,M_{\vec{i}})$ are estimators for $f(\mathcal{E},\bigotimes_{j=1}^Q|0\rangle\langle0|,M_{\vec{i}})$.
    \end{enumerate}
    \item \emph{Estimate Noiseless Value:} 
    \begin{enumerate}
        \item For each non-zero $c_{\vec{i}}$, estimate the noiseless expectation value $\Tr[\rho P_{\vec{i}}]$ as
        \begin{align}
            \widehat{\Tr[\rho P_{\vec{i}}]} =\frac{\hat{f}(\mathcal{E},\rho,P_{\vec{i}})}{\hat{f}(\mathcal{E},\bigotimes_{j=1}^Q|0\rangle\langle0|,M_{\vec{i}})}.
        \end{align}
        \item Estimate the noiseless expectation value $\Tr[\rho O]$ as
        \begin{align}
            \widehat{\Tr[\rho O]} = \sum_{\vec{i}} c_{\vec{i}}\widehat{\Tr[\rho P_{\vec{i}}]}.
        \end{align}
    \end{enumerate}
\end{enumerate}

To make sense of this procedure it is important to recognize that, by definition, $\bigotimes_{j=1}^Q|0\rangle$ is a $+1$ eigenstate of all $M_{\vec{i}}$'s. We therefore find that
\begin{align}
    \hat{f}(\mathcal{E},\bigotimes_{j=1}^Q|0\rangle\langle0|,M_{\vec{i}})=\frac{1}{2^Q}\Tr[M_{\vec{i}}\mathcal{E}(M_{\vec{i}})]
\end{align}
is the suppression factor on the expectation value in Equation~\eqref{eq:suppression}.

\subsection{Tensor Product Approximation}
In the case of working with a tensor product approximation to the noise model the procedure is very similar to the above. The difference is that the step labeled \emph{Noise Suppression Estimation} (Step~\ref{step:suppression}) is replaced with
\begin{enumerate}
    \item[2'.] \emph{Tensor Product Noise Suppression Estimation}
    \begin{enumerate}
        \item Prepare and measure the all zeros state $\bigotimes_{j=1}^Q|0\rangle\langle0|$ a number of times $N_c$ with randomly drawn unitary rotations immediately prior to the measurement.
        \item For each qubit index $\ell$ for which there is at least one non-zero $c_{\vec{i}}$ with $i_{\ell}\neq0$, record the sample mean of the noisy, single shot estimators of the expectation value of $\sigma_z$ on that qubit as
        \begin{align}
            \hat{f}(\mathcal{E},\bigotimes_{j=1}^Q|0\rangle\langle0|,M_{\hat{\ell}}) = \frac{1}{N_{c}}\sum_{a=1}^{N_c}3 m_{\ell,a} \vec{n}_{\ell,a}\cdot\hat{z}.
        \end{align} 
        Here $M_{\hat{\ell}} = \I^{\otimes(\ell-1)}\otimes\sigma_z\otimes\I^{\otimes(Q-\ell)}$.
        As above, these $\hat{f}(\mathcal{E},\bigotimes_{j=1}^Q|0\rangle\langle0|,M_{\hat{\ell}})$ are estimators for $f(\mathcal{E},\bigotimes_{j=1}^Q|0\rangle\langle0|,M_{\hat{\ell}})$. For a generic Pauli string $P_{\vec{i}}$ we then construct an estimate for 
        \begin{align}
            \hat{f}'(\mathcal{E},\bigotimes_{j=1}^Q|0\rangle\langle0|,M_{\vec{i}})= \prod_{\{\ell \big|i_{\ell}\neq0\}}\hat{f}(\mathcal{E},\bigotimes_{j=1}^Q|0\rangle\langle0|,M_{\hat{\ell}}).
        \end{align}
        Here $\hat{f}'(\mathcal{E},\bigotimes_{j=1}^Q|0\rangle\langle0|,M_{\vec{i}})$ is the tensor product estimator of the suppression factor.
    \end{enumerate}
\end{enumerate}

Note that when working in a setting where the tensor product approximation is sufficient, we only need to estimate a single suppression value for each qubit. While the measurement procedure is the same whether or not the tensor product approximation is taken, the computation of the estimated suppression factors can be done once for all observables. More generally, we would need to separately estimate a suppression value for each set of Pauli strings with the same placement of identity operators.

\subsection{Resource Scaling}
 Here we are examine the shot cost of using means as expectations, but it is common in the classical shadow literature to instead work with a median-of-means estimator~\cite{huang2020predicting}. For a nice treatment of the shot cost using the median-of-means estimator, see~\cite{chen2021robust}.

In order to estimate the resources required to mitigate the expectation value of $P_{\vec{i}}$ to a precision $\epsilon$, we first estimate the variance of $\widehat{\Tr[\rho P_{\vec{i}}]}$. Noting that both $\hat{f}\left(\mathcal{E},\rho,P_{\vec{i}}\right)$ and $\hat{f}\left(\mathcal{E},\bigotimes_{j=1}^Q|0\rangle\langle0|,M_{\vec{i}}\right)$ will usually be estimated with at least thousands of samples, one can use the central limit theorem to describe the distribution of sampled mean values of these estimators as approximately normal. With this in mind, we follow~\cite{diaz2013existence} and use a second-order approximation to the variance:
\begin{align}
    \var\left(\widehat{\Tr[\rho P_{\vec{i}}]}\right) \approx & \frac{\mathbbm{E}\left[\hat{f}\left(\mathcal{E},\rho,P_{\vec{i}}\right)\right]^2}{\mathbbm{E}\left[\hat{f}\left(\mathcal{E},\bigotimes_{j=1}^Q|0\rangle\langle0|,M_{\vec{i}}\right)\right]^2}\nonumber \\
    &\cdot\Bigg(\frac{\var\left(\hat{f}\left(\mathcal{E},\rho,P_{\vec{i}}\right)\right)}{\mathbbm{E}\left[\hat{f}\left(\mathcal{E},\rho,P_{\vec{i}}\right)\right]^2}\nonumber\\
    &\:+\frac{\var\left(\hat{f}\left(\mathcal{E},\bigotimes_{j=1}^Q|0\rangle\langle0|,M_{\vec{i}}\right)\right)}{\mathbbm{E}\left[\hat{f}\left(\mathcal{E},\bigotimes_{j=1}^Q|0\rangle\langle0|,M_{\vec{i}}\right)\right]^2}\Bigg).
\end{align}

As the estimators we take the expectation value of are unbiased, we can replace these expectation values with the quantity to be estimated. Bounding the variance of these estimators with their seminorm bounds, we then have
\begin{align}\label{eq:ratio var bound}
    \var\left(\widehat{\Tr[\rho P_{\vec{i}}]}\right) \lesssim & \frac{1}{N_{\rho}}\Bigg(\frac{\|P_{\vec{i}}\|^2}{\left(\frac{1}{2^Q}\Tr[M_{\vec{i}}\mathcal{E}(M_{\vec{i}})]\right)^2}\nonumber\\
    &\:+\frac{\Tr[\rho P_{\vec{i}}]^2\|M_{\vec{i}}\|^2}{N_c/N_{\rho}\left(\frac{1}{2^Q}\Tr[M_{\vec{i}}\mathcal{E}(M_{\vec{i}})]\right)^2}\Bigg).
\end{align}

Setting the right side of Equation~\eqref{eq:ratio var bound} equal to the desired variance $\epsilon^2$ and denoting the ratio of the number shots spent on estimating the calibration ($N_c$) to the number of shots used to estimate the desired expectation value ($N_{\rho}$) as $b=N_c/N_{\rho}$, we find
\begin{align}
    N_{\rho}  = \frac{1}{\epsilon^2}&\Bigg(\frac{\|P_{\vec{i}}\|^2}{\left(\frac{1}{2^Q}\Tr[M_{\vec{i}}\mathcal{E}(M_{\vec{i}})]\right)^2}\nonumber\\
    &\:+\frac{\Tr[\rho P_{\vec{i}}]^2\|M_{\vec{i}}\|^2}{b\left(\frac{1}{2^Q}\Tr[M_{\vec{i}}\mathcal{E}(M_{\vec{i}})]\right)^2}\Bigg).
\end{align}
To find the optimal choice of $b$, we seek to minimize the overall shot cost $N_{\textrm{total}}= N_{\rho}(1+b)$. The minimum value for $N_{\textrm{total}}$ occurs at
\begin{align}
    b = \frac{\Tr[\rho P_{\vec{i}}]^2\|M_{\vec{i}}\|^2}{\|P_{\vec{i}}\|^2},
\end{align}
which gives
\begin{align}
    N_{\textrm{total}} =& \frac{2\|P_{\vec{i}}\|^2}{\epsilon^2\left(\frac{1}{2^Q}\Tr[M_{\vec{i}}\mathcal{E}(M_{\vec{i}})]\right)^2}\nonumber \\
    &\cdot \left(1+\frac{\Tr[\rho P_{\vec{i}}]^2\|M_{\vec{i}}\|^2}{\|P_{\vec{i}}\|^2} \right)
\end{align}
We note that this results in similar scaling to that found with the median-of-means estimator~\cite{chen2021robust}.

Finally, we remark that working with the pole-concentrated estimator gives the same result with the seminorm $\|\cdot\|$ replaced with $\|\cdot\|'$. 

Using the approximate state framework in this manner allows one to mitigate readout errors with resource scaling that depends only on the locality of the operators of interest, not the overall size of the state. We also note that, as a practical matter, if desired level of precision is such that hardware drift is not an impediment, it may be enough to only collect the $N_c$ calibration shots once as a preparatory step before performing the desired computations.

%%%%%%%%%%%%%%%%%%%%%%%%%%%%%%%%%%%%%%%%%%%%%%
%                                                HARDWARE IMPLEMENTATION                                             %
%%%%%%%%%%%%%%%%%%%%%%%%%%%%%%%%%%%%%%%%%%%%%%
%\maketitle
\section{Hardware Implementation} \label{sec:hardware}
Here we discuss how we efficiently implement the approximate state framework on the Rigetti hardware stack.

\subsection{Pole-Concentrated Sampling}
As discussed in section~\ref{sec:non_uniform}, the pole-concentrated sampling technique can be readily mapped to a fast implementation within the live digital signal processors or Field Programmable Gate Arrays (FPGAs) which are commonly now found in superconducting circuit control systems.  

Note that an arbitrary single-qubit gate can be mapped to a standard sequence of five gates comprising of variable angle phase gates and quarter-cycle rotations around another axis: 

\begin{equation}
    U(\alpha, \beta, \gamma) = \mathcal{R}_Z(\gamma)\mathcal{R}_X\left(\frac{\pi}{2}\right)\mathcal{R}_Z(\beta)\mathcal{R}_X\left(\frac{\pi}{2}\right)\mathcal{R}_Z(\alpha),
\end{equation}
where $\mathcal{R}_X(\cdot)$ and $\mathcal{R}_Z(\cdot)$ are rotations generated by the $X$ and $Z$ Pauli matrices.

The final degree of freedom (here in the angle $\lambda$) is unnecessary as measurements in the Z-basis performed directly following such a rotation will be unaffected by it.

This decomposition is useful, as implementation of phase gates can be performed in a 'virtual' manner, wherein the inverse phase shift is performed on the XY-control signal line itself rather than the physical qubit, representing a change of basis as an alternative to a physical change of state for the qubit~\cite{mckay2017efficient}.  The $\mathcal{R}_X(\pi/2)$ gate is implemented as a microwave control signal which is previously calibrated.

Upon each repetition of a program making use of random phases before measurement, the sequence of two interleaved phase shifts and microwave pulses must therefore be performed before readout.  Such operations are a part of the standard processing chain as they are required for all basic operations.  Additionally, generation of pseudo-random numbers on the fly via an XOR LFSR (Linear Feedback Shift Register) protocol, which is both fast and easily implemented on FPGAs, can be used as the source of phase shift values.  This pseudo-random sequence is seeded differently for each qubit in each execution batch.  The same sequence can then also be generated on the conventional CPU where post-processing is to be performed.  This allows for execution of a circuit a large number of times with different randomly generated readout bases, without the need for re-compilation or multiple execution requests.

\subsection{Tetrahedral Sampling}
Sampling from a discrete group of rotations, such as the tetrahedral group, can be performed on hardware almost as quickly as the pole-concentrated sampling. The procedure is the same as for the case of the pole-concentrated sampling except that the pseudo-random numbers generated by the XOR LFSR protocol are passed to predetermined lookup tables of phases stored on the FPGAs.

\subsection{Handling Drift}\label{sec:drift}
Changes in the noise characteristics of a quantum device over time, called drift, pose a problem for all error mitigation techniques. Essentially, if the calibration for the error mitigation is no longer accurate by the time of the collection of the data of interest, the results of the mitigation can contain a significant bias. 

To reduce such biases, it is best to gather the calibration data as close in time as possible to gathering the data from the circuit of interest. We have empirically found collecting the two different data sets in an interleaved fashion is very important to achieving a high precision mitigation with this method. This interleaved execution is implemented by batching the shots used for both the desired data and the calibration data and then executing the batches in an alternating (interleaved) order.

%%%%%%%%%%%%%%%%%%%%%%%%%%%%%%%%%%%%%%%%%%%%%%
%                                                EXPERIMENTAL RESULTS                                                %
%%%%%%%%%%%%%%%%%%%%%%%%%%%%%%%%%%%%%%%%%%%%%%
%\maketitle
\section{Experimental Results} \label{Experimental Results}

\subsection{Suppression of Correlated Errors}\label{sec:correlated error demo}
As mentioned in Section~\ref{sec:The Randomised Readout Channel}, the high degree of symmetrization of the effective error channel that results from the approximate state procedure can lead to a significant reduction of correlated errors that are asymmetric. Here we demonstrate that by measuring all pairwise Pearson correlation coefficients between qubits measured in the $|0\rangle^{\otimes N}$ state on the Rigetti Aspen-11 quantum processing unit (QPU). As this state gives rise to no correlation between qubits, these correlation coefficients can be interpreted as correlations induced by errors. Further, given that the state preparation typically has a much higher fidelity than readout, to a good approximation these correlations are induced by readout errors.

\begin{figure}
    \centering
    \includegraphics[width=1.0\columnwidth]{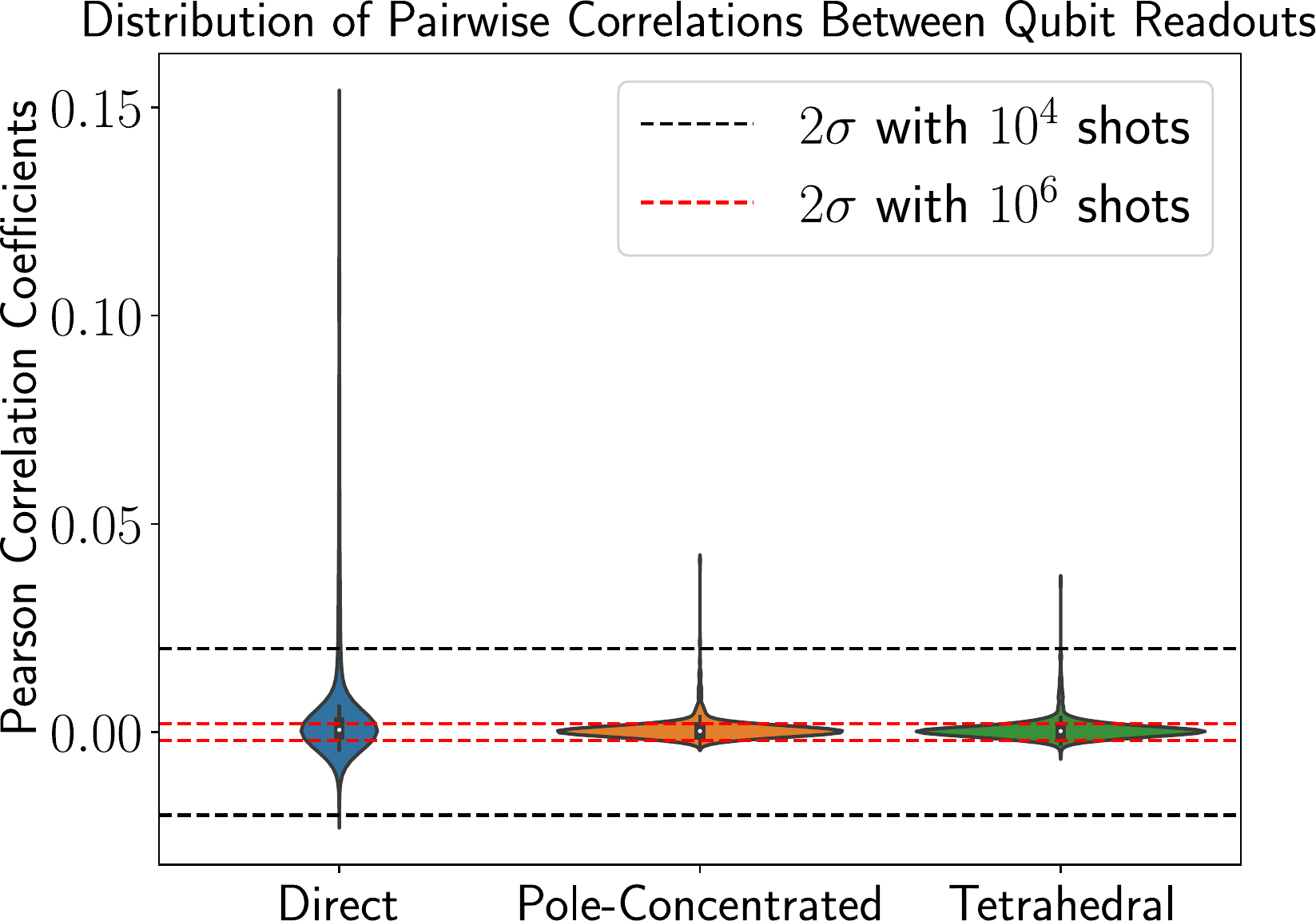}
    \caption{The distribution of correlations between all pairs of qubits on Aspen-11. These correlations are computed using the Direct, Pole-Concentrated, and Tetrahedral methods discussed in the main text. These correlations are measured with $10^6$ shots and the dashed red lines are the $2\sigma$ bounds expected for a measurement of a true $0$ correlation, while the dashed black line shows the same for using $10^4$ shots.}
    \label{fig:corr}
\end{figure}

Figure~\ref{fig:corr} shows the distribution of readout-induced correlations over the set of all pairs of qubits on the Aspen-11 QPU for three different readout strategies. These strategies are directly reading out the state without any randomisation (Direct), using the pole-concentrated randomisation (Pole-Concentrated), and finally using the tetrahedral randomisation (Tetrahedral). We have included the $2-\sigma$ detection bounds for correlations with experiments using $10^4$ and $10^6$ shots as dashed lines. These measurements were taken using $10^6$ shots for each measurement method.

For all cases, the readout induced correlations are small enough that, outside of a few outliers, these correlations would not even be detectable in experiments using $~10^4$ shots.

We find that while most of the readout induced correlations in the Direct method are small, both the Pole-Concentrated and Tetrahedral methods exhibit significantly less correlation between qubits.

For the Direct, Pole-Concentrated, and Tetrahedral methods collecting this data took $63$ seconds, $87$ seconds, and $84$ seconds, respectively.

\subsection{Improving the Results of Quantum Simulations of Plasma Physics}
\begin{figure}[t!]
    \centering
    \includegraphics[width=1.0\columnwidth]{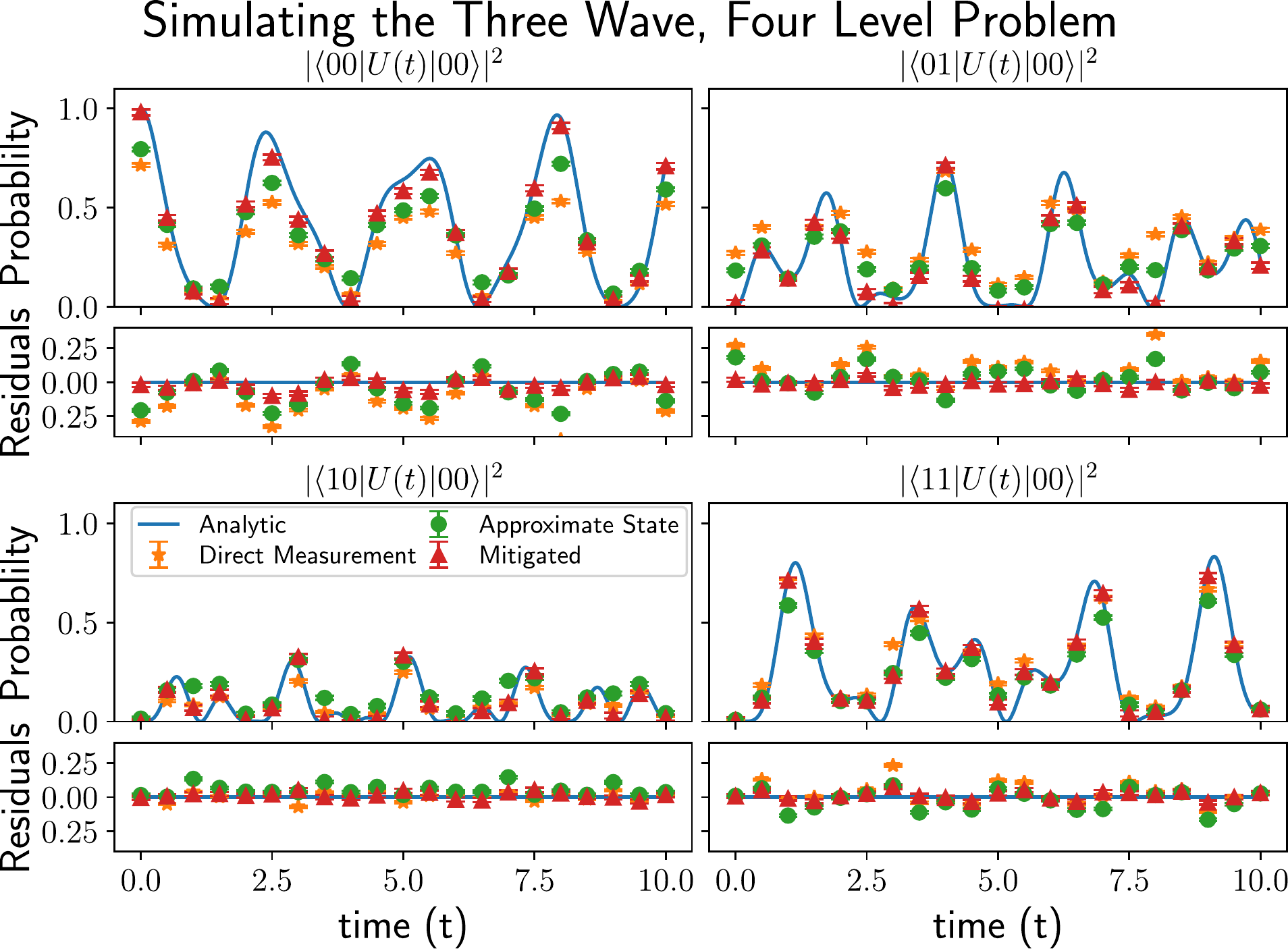}
    \caption{The probability of finding given bit strings when measuring the simulated state, as a function of simulation time. The readout mitigated estimates (red triangles) show the simulation following the analytic solution much more closely than it appears without readout mitigation.}
    \label{fig:three_wave}
\end{figure}
As a demonstration of this method, we examine applying the approximate state readout mitigation strategy to the simulation of the four-level, three-wave problem. The three-wave problem is a cubic interaction problem that represents the lowest-order, nonlinear coupling in quantum optics and plasma dynamics~\cite{shi2021simulating}. This problem corresponds to the following Hamiltonian:
\begin{equation}
    H = ig a^{\dagger}_1a_2a_3 - i g^*a_1a^{\dagger}_2a^{\dagger}_3,
\end{equation}
with symmetries allowing for the reduction of the dynamical space as discussed in~\cite{shi2021simulating}. Specifically, the operators
\begin{equation}
    S_2 = a^{\dagger}_1a_1 + a^{\dagger}_2a_2
\end{equation}
and 
\begin{equation}
    S_3 = a^{\dagger}_1a_1 + a^{\dagger}_3a_3
\end{equation}
commute with this Hamiltonian and are thus conserved. Fixing the eigenvalues of both $S_2$ and $S_3$ to be $3$, we are left with a four level system that can then be simulated with two qubits~\cite{shi2021simulating}.

\begin{figure}[t!]
    \centering
    \includegraphics[width=1.0\columnwidth]{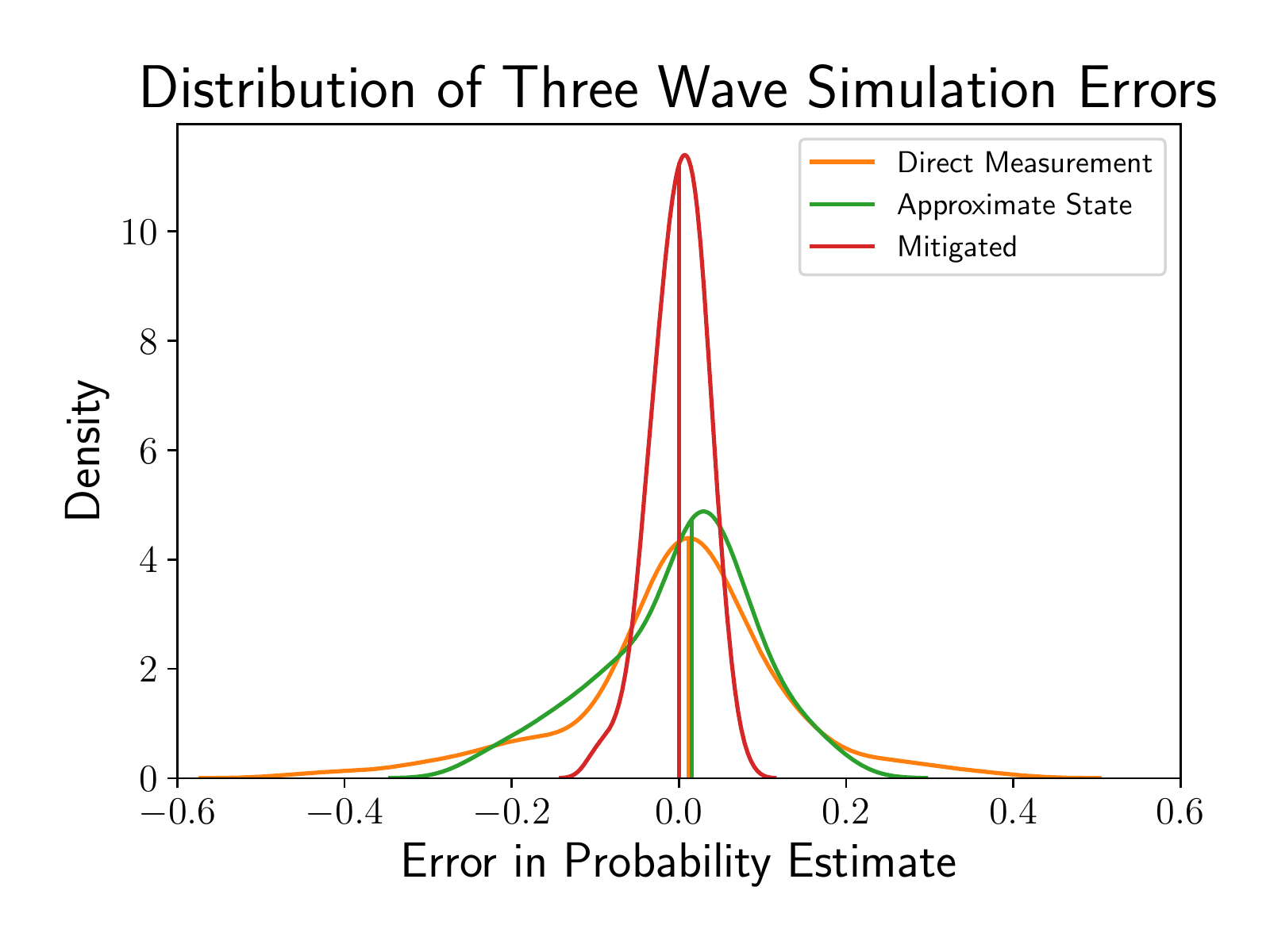}
    \caption{The distribution of errors in the simulation of the three wave problem. These kernel density estimation curves show a smoothed distribution reflecting the absolute errors of the data points in Figure~\ref{fig:three_wave}.}
    \label{fig:three_wave_errs}
\end{figure}

In Figure~\ref{fig:three_wave} we show the probability distribution of finding this simulated system in any of the computational basis states as a function of time, starting from the $|00\rangle$ initial state. As can be seen in Figure~\ref{fig:three_wave_errs}, the readout mitigation substantially decreases the errors in the estimated probabilites. Here the direct estimation is done with $10^4$ shots per data point, taking 24 seconds to collect the data for all points shown. The approximate state measurements were were sampled with the tetrahedral randomisation. These measurements were taken with $10^5$ shots per data point and $9.5\cdot10^5$ shots used for the calibration of each data point. The combined sampling and calibration for the approximate state approaches took $13$ min $26$ seconds.

For larger systems this application of the approximate state framework will not scale well as the operators in question are projectors on the whole system, incurring an exponential overhead. However, this framework would be useful for studying the behaviors of subsystems and local properties.

%%%%%%%%%%%%%%%%%%%%%%%%%%%%%%%%%%%%%%%%%%%%%%
%                                                CONCLUSIONS                                                          %
%%%%%%%%%%%%%%%%%%%%%%%%%%%%%%%%%%%%%%%%%%%%%%
%\maketitle
\section{Conclusions} \label{Conclusions}
%Basic Idea
As near-term applications of quantum computers must contend with significant errors, these applications will need to be designed with errors in mind. Readout errors in particular can be a significant hurdle for achieving useful results from a method, making strategies to mitigate these errors crucial. In this manuscript we have introduced readout error mitigation in the approximate state framework~\cite{paini2019approximate}, providing a way to remove the impact of readout errors from the approximate state estimators. As many near-term applications will require the use of estimators like the ones the approximate state framework provides, this error mitigation may prove to be a useful subroutine across a variety of algorithms.

%Theoretical Summary
Much like in the case of the closely related classical shadow framework~\cite{huang2020predicting}, the approximate state framework allows for the efficient estimation of many properties of quantum states from randomised measurements. This randomisation of the measurements, similar to twirling a gate, symmetrises the readout error~\cite{chen2021robust}. We have applied these ideas to various sampling strategies for the approximate state framework and explicitly computed the resulting effect of general error channels. We found that all expectation values are suppressed by a constant factor that differs from what might have been expected from the previously made analogy with gate twirling.

%Rigetti Hardware Acceleration
Using standard approaches with modern hardware, applying either the approximate state or classical shadow frameworks has been prohibitively slow to execute for practical purposes. The bottleneck in making use of these techniques comes from the need to change the final gates implemented after each measurement, introducing a significant amount of latency with each sample. Our implementation of a pseudo-random number generation on the FPGAs in the Rigetti control systems allows us to randomly sample different measurements efficiently, removing this latency. Specifically, in our hardware demonstration we have shown that we can sample and process one million shots, each corresponding to a random measurement in the approximate state framework, in less than a minute and a half. We estimate that performing the same operation without the advantage of our FPGA randomisation would have taken roughly seventeen hours, making these advances necessary for the practical application of the approximate state framework.

%Demonstration Commentary
Aside from being fast, we have also demonstrated that on current Rigetti hardware the small correlations that can be introduced during readout are essentially eliminated when using the approximate state technique. With direct measurements we found small but detectable error-induced correlations from the readout process on the Aspen-11 chip, but utilising the approximate state formalism essentially eliminated those correlations. As a consequence, it follows that the tensor product approximation to the effective readout channel in the approximate state form is justified and thus complex estimators can be mitigated with a simple product of single qubit terms.

%Removal of readout allows to see how bad other errors are
For the case of mitigating the readout errors in a plasma physics simulation, we found the approximate state mitigation approach to be successful. For this application much of the error came from readout, meaning that the state achieved with digital quantum simulation was more faithful than it might have seemed. Given this result, the efficient and accurate readout mitigation our method achieves may be useful not only for improving the end results of a computation but also as a way to separate the impact of different types of error on quantum computers performance for a variety of near-term applications.

%Combination with other mitigation strategies.
Finally, we remark that early practical applications of quantum computers will likely require error mitigation of gates as well as of readout. As the formalism presented here provides readout-mitigated expectation values, it would naturally lend itself to the various expectation value error mitigation schemes such as zero noise extrapolation (ZNE) and Clifford data regression (CDR)~\cite{temme2017error,endo2018practical, czarnik2021error, lowe2021unified}. Furthermore, the hardware acceleration we have developed could be extended to also provide the random sampling for randomised compilation, speeding up that method as well. 

%%%%%%%%%%%%%%%%%%%%%%%%%%%%%%%%%%%%%%%%%%%%%%
%                                                      ACKNOWLEDGEMENTS                                                  %
%%%%%%%%%%%%%%%%%%%%%%%%%%%%%%%%%%%%%%%%%%%%%%

\section{Acknowledgments}

The majority of this project was supported by Innovate UK grant 10001505. The methods for handling drift in Section~\ref{sec:drift} and the experimental demonstrations in Section~\ref{Experimental Results} are based upon work supported by the U.S. Department of Energy, Office of Science, SC-1
U.S. Department of Energy
1000 Independence Avenue, S.W.
Washington DC 20585,
under Award Number(s) DE-SC0021661. Any opinions, findings, and conclusions or recommendations expressed in this publication are those of the author(s) and do not necessarily reflect the views of the U.S. Department of Energy.

Additionally, the authors would like to thank Bram Evert and the rest of the Rigetti team for helpful discussions and suggestions. We also thank Amy Brown, Daniel Lidar, Joseph Ilon, Max Porter, Phattharaporn Singkanipa, Shi Yuan, Vasily Geyko, Vinay Tripathi and Yujin Cho for discussions related to the simulation of the three wave plasma physics problem. Finally, we thank Yujin Cho for the use of code that generated the unitaries for that three wave problem simulation. 
%%%%%%%%%%%%%%%%%%%%%%%%%%%%%%%%%%%%%%%%%%%%%%
%                                                           BIBLIOGRAPHY                                                           %
%%%%%%%%%%%%%%%%%%%%%%%%%%%%%%%%%%%%%%%%%%%%%%

\bibliography{rigetti}

%%%%%%%%%%%%%%%%%%%%%%%%%%%%%%%%%%%%%%%%%%%%%%
%                                                           APPENDICES                                                  %
%%%%%%%%%%%%%%%%%%%%%%%%%%%%%%%%%%%%%%%%%%%%%%
%\maketitle
\onecolumngrid
\clearpage
\appendix
%%%%%%%%%%%%%%%%%%%%%%%%%%%%%%%%%%%%%%%%%%%%%%
%                                  APPENDIX F - VARIANCE BOUNDS FOR DIFFERENT SAMPLING SETS					    	 %
%%%%%%%%%%%%%%%%%%%%%%%%%%%%%%%%%%%%%%%%%%%%%%

\section{Alternative Sampling Sets} \label{App Variance Bounds for Different Sampling Sets}

\subsection{Sampling t-designs}\label{app:t-design}

In the main text we note that the approximate state formalism's uniform sampling of Bloch sphere measurement directions can be replaced by sampling from a t$\geq$2-design. Here we demonstrate the equivalence of the second moments of single-qubit operators when sampling either directions $\vec{n}$ or unitaries from $SU(2)$.

To bound the variance of the single-qubit estimators we consider the expectation value of~\cite{paini2021estimating}: 
\begin{align}
    \langle R_1[\sigma_r](m,\Vec{n}) R_1[\sigma_s](m,\Vec{n})\rangle =& \int \frac{d\hat{n}}{4\pi} \sum_{m=\in\{-1,1\}} p(m,\hat{n})\nonumber \\
    & \quad R_1[\sigma_r](m,\hat{n}) R_1[\sigma_s](m,\hat{n})\nonumber \\
    =& 9 \int\frac{d\hat{n}}{4\pi} n_r n_s\nonumber\\
    =& 3\delta_{r,s}
\end{align}

Now we will switch from integrating uniformly over directions on the unit sphere to the equivalent integral over the Haar distribution on $SU(2)$. 
\begin{align}
    \langle R_1[\sigma_r](m,\Vec{n}) R_1[\sigma_s](m,\Vec{n})\rangle =& \frac{9}{4}\int dU \Tr\left[U^\dagger \sigma_Z U \sigma_r\right]\Tr\left[U^\dagger \sigma_Z U \sigma_s\right]\nonumber \\
    =& 9\int dU \Tr\left[U^{\dagger^{\otimes 2}} \sigma_Z^{\otimes 2}  U^{\otimes 2}  (\sigma_r \otimes \sigma_s)\right]\nonumber \\ 
    =& \frac{3}{2} \Tr[\sigma_r \sigma_s]\nonumber\\
    =& 3 \delta_{r,s}
\end{align}
Here we have used the Haar integration identity from Equation 10 of~\cite{puchala2017symbolic}. Note that since we have terms that look like $U^{\otimes 2}$, but no higher tensor powers, averaging over any ensemble of single-qubit unitaries that is at least a 2-design will yield the same result by definition~\cite{dankert2009exact}.

The result that the variance and covariances of multi-qubit estimators are equivalent for both sampling methods follows directly from~\cite{paini2021estimating} by making this same mapping from integrating over Bloch spheres to integrating over single-qubit Haar measures.

\subsection{Pole-Concentrated Sampling}\label{app:pole_concentrated}
As discussed in Section~\ref{sec:non_uniform} of the main text, if we sample the measurement directions for each qubit non-uniformly on the Bloch sphere we can reduce the variance for some operators at the cost of increasing it for others. Here we go into the details of how this could be done for a simple example resulting in decreased variance for tensor products of Pauli $Z$ operators and identities.

In order to do this without biasing the expectation value estimators, in Section~\ref{sec:non_uniform} we defined a new single-qubit kernel operator $K'_1(m,\vec{n})$ and associated single-qubit expectation value estimator $R'_1$ from the spherically symmetric kernel operator $K_1(m,\vec{n})$. Repeating those definitions here we have:
\begin{align}
    K'_1(m,\vec{n}(\theta,\phi)) \equiv \frac{\pi}{2}\textrm{sin}(\theta)K_1(m,\vec{n}(\theta,\phi)),
\end{align}
and
\begin{align}
    R'_1[O] =& \frac{1}{N}\sum_{j=1}^N \Tr\left[K_1'(m_j,\vec{n}_j) O\right]\nonumber\\
    =&\frac{\pi}{2N}\sum_{j=1}^N \textrm{sin}(\theta)\Tr\left[K_1(m_j,\vec{n}_j) O\right].
\end{align}
These were defined so that the expected value of this estimator is the same as for spherically symmetric sampling, that is $\langle R'_1[O]\rangle=\langle R_1 [O]\rangle=\Tr[\rho O]$. As noted in the main text, higher moments of the different estimators do not agree in general. Here we will discuss the variance of this pole-concentrated estimator.

\subsubsection{Expectation Values of Products of Pairs of Single-Qubit Pauli Estimators}
For the pole concentrated sampling, we need to examine $\langle R_1'[\sigma_r](m,\Vec{n}) R_1'[\sigma_s](m,\Vec{n})\rangle$ for different values of $r$ and $s$. 

For both $r$ and $s$ in $\{X,Y,Z\}$ (i.e. not $\I$) we have that
\begin{align}
    \langle R_1'[\sigma_r](m,\Vec{n}) R_1'[\sigma_s](m,\Vec{n})\rangle =& \int_0^{2\pi}\frac{d\phi}{2\pi}\int_0^{\pi}\frac{d\theta}{\pi} \sum_{m \in \{-1,1\}}\left(\frac{3\pi}{2}\right)^2\textrm{sin}^2(\theta)p(m,\Vec{n})n_{r}n_{s}\nonumber\\
    =& \frac{9}{8}\int_0^{2\pi}d\phi\int_0^{\pi} d\theta\sum_{m \in \{-1,1\}} \textrm{sin}^2(\theta)p(m,\Vec{n})n_{r}n_{s}\nonumber\\
    =& \frac{9}{8}\int_0^{2\pi}d\phi\int_0^{\pi}d\theta\,\textrm{sin}^2(\theta) n_{r}n_{s}.
\end{align}

First, if $r \ne s$ and $r,\,s\in\{X,Y,Z\}$ we have 
\begin{equation}
    \langle R_1'[\sigma_r](m,\Vec{n}) R_1'[\sigma_s](m,\Vec{n})\rangle = 0. \nonumber
\end{equation}
Unlike the spherically symmetric case, however, the $Z$ direction is now different from the $X$ and $Y$ directions. For $r = s = Z$ we have
\begin{align}
    \langle R_1'[\sigma_Z](m,\Vec{n})^2\rangle
    =& \frac{9 \pi}{4}\int_0^{\pi}d\theta\,\textrm{sin}^2(\theta) \textrm{cos}^2(\theta)\nonumber \\
    =& \frac{9 \pi^2}{32} \nonumber \\
    \ge & \textrm{Var}[ R_1'[\sigma_Z]].  \nonumber
\end{align}
Therefore, sampling $\theta$ uniformly from the interval $[0,\pi]$ results in the variance of the single-shot estimator of $\langle\sigma_Z\rangle$ being bounded by $\frac{9 \pi^2}{32} \sim 2.78$. Note that this bound is slightly smaller than for the spherically symmetric case.

Considering instead the case of $r = s = X$ we have
\begin{align}
    \langle R_1'[\sigma_X](m,\Vec{n})^2\rangle
    =& \frac{9}{8}\int_0^{\pi}d\theta\,\textrm{sin}^4(\theta) \int_0^{2\pi}\textrm{cos}^2(\phi) d \phi\nonumber \\
    =& \frac{27 \pi^2}{64} \nonumber \\
    \ge & \textrm{Var}[ R_1'[\sigma_X]]. \nonumber
\end{align}
Similarly, for the case of $r = s = Y$ we have
\begin{align}
    \langle R_1'[\sigma_Y](m,\Vec{n})^2\rangle
    =& \frac{9}{8}\int_0^{\pi}d\theta\,\textrm{sin}^4(\theta) \int_0^{2\pi}\textrm{sin}^2(\phi) d \phi\nonumber \\
    =& \frac{27 \pi^2}{64} \nonumber \\
    \ge & \textrm{Var}[ R_1'[\sigma_Y]].
\end{align}
We then have that sampling $\theta$ uniformly from the interval $[0,\pi]$ results in the single-shot estimator of $\langle\sigma_X\rangle$ or $\langle\sigma_Y\rangle$ being bounded by $\frac{27 \pi^2}{64} \sim 4.16$. We note that this bound is larger than for the spherically symmetric case. However, if this sampling strategy can be implemented at least $\sim 1.4$ times faster than the symmetric one, it will still result in a lower overall variance bound with the same time investment for these single-qubit operators.

In addition to the above cases, with this estimator the identity components are also random variables and so we need to know their variance. We therefore also consider the case of $r = s = \I$:
\begin{align}
    \langle R_1'[\I](m,\Vec{n})^2\rangle
    =& \frac{1}{8}\int_0^{\pi}d\theta\,\textrm{sin}^2(\theta) \int_0^{2\pi}d \phi\nonumber \\
    =& \frac{\pi^2}{8}  \nonumber
\end{align}
Note that, unlike the other cases, for the identity operator the mean of the estimator is exactly $1$ for any state, which means that the variance is precisely
\begin{equation}
    \textrm{Var}\left(R_1'[\I]\right)=\frac{\pi^2}{8}-1  \nonumber
\end{equation}
and we have no need for a bound.

Finally, let's consider $r = \I$ and $s \neq \I$.
\begin{align}
    \langle R_1'[\I](m,\Vec{n}) R_1'[\sigma_s](m,\Vec{n})\rangle =& \int_0^{2\pi}\frac{d\phi}{2\pi}\int_0^{\pi}\frac{d\theta}{\pi} \sum_{m \in \{-1,1\}} 3\left(\frac{\pi}{2}\right)^2m\,\textrm{sin}^2(\theta)p(m,\Vec{n})n_{s}\nonumber\\
    =& \frac{3}{8}\int_0^{2\pi}d\phi\,\textrm{sin}^2(\theta)\int_0^{\pi}d\theta n_{s} \sum_{m \in \{-1,1\}}mp(m,\Vec{n})\nonumber\\
    =& \frac{3}{8}\int_0^{2\pi}d\theta\, \textrm{sin}^2(\theta)\int_0^{\pi}d\phi n_{s}\sum_\gamma n_{\gamma} \Tr[\rho\sigma_\gamma].
\end{align}\
In the last line here we have identified $\sum_{m \in \{-1,1\}}mp(m,\Vec{n})$ as $\Tr[\rho \vec{\sigma}\cdot \vec{n}]$ since it is an expectation value, then expanded the inner product. We then re-write this as
\begin{align}
    \langle R_1'[\I](m,\Vec{n}) R_1'[\sigma_s](m,\Vec{n})\rangle =& \frac{3}{8}\sum_\gamma  \Tr[\rho\sigma_\gamma]\nonumber\int_0^{\pi}d\theta\,\textrm{sin}^2(\theta)\int_0^{2\pi}d\phi n_{s}n_{\gamma}. \nonumber
\end{align}
This is the same integral we had for the case where both $r$ and $s$ were not the identity, so we can simplify by enforcing $s =\gamma$:
\begin{align}
    \langle R_1'[\I](m,\Vec{n}) R_1'[\sigma_s](m,\Vec{n})\rangle =& \frac{3}{8} \Tr[\rho\sigma_s] \int_0^{\pi}d\theta\,\textrm{sin}^2(\theta)\int_0^{2\pi}d\phi n_{s}^2.
\end{align}
For $s = Z$ we then have:
\begin{align}
    \langle R_1'[\I](m,\Vec{n}) R_1'[\sigma_Z](m,\Vec{n})\rangle =&\frac{3 \pi}{4}\int_0^{\pi}d\theta\,\textrm{sin}^2(\theta) \textrm{cos}^2(\theta)\Tr[\rho Z]\nonumber\\
    =& \frac{3 \pi^2}{32}\Tr[\rho Z].
\end{align}
Similarly we find

\begin{align}
    \langle R_1'[\I](m,\Vec{n}) R_1'[\sigma_X](m,\Vec{n})\rangle =&\frac{3 \pi^2}{8}\int_0^{\pi}d\theta\,\textrm{sin}^4(\theta) \int_0^{2\pi}\textrm{cos}^2(\phi) d \phi\Tr[\rho X]\nonumber\\
    =& \frac{9 \pi^2}{64}\Tr[\rho X]
\end{align}
and 
\begin{align}
    \langle R_1'[\I](m,\Vec{n}) R_1'[\sigma_Y](m,\Vec{n})\rangle =&\frac{3 \pi^2}{8}\int_0^{\pi}d\theta\,\textrm{sin}^4(\theta) \int_0^{2\pi}\textrm{sin}^2(\phi) d \phi \Tr[\rho Y]\nonumber\\
    =& \frac{9 \pi^2}{64}\Tr[\rho Y].  \nonumber
\end{align}
for $s=X$ and $s=Y$, respectively.

\subsubsection{Expectation Values of Products of Pairs of Multi-Qubit Pauli Estimators}

We now extend this discussion to the case of multiple qubit products such as $\langle R'[ P_{\vec{i}}] R'[ P_{\vec{j}}]\rangle$. We can expand this type of product as:
\begin{align}
    \langle R'[ P_{\vec{i}}] R'[ P_{\vec{k}}]\rangle =& \prod_{j\in \mathcal{S}} \Bigg(\int \frac{d\phi_j}{2\pi}\int \frac{d \theta_j}{\pi}\sum_{m_j \in \{-1,1\}} \left(\frac{3\pi}{2}\right)^2\textrm{sin}^2(\theta_j)n_{i_j}n_{k_j} \Bigg)\nonumber\\
    &\cdot \prod_{j\in \mathcal{S}'} \Bigg(\int \frac{d\phi_j}{2\pi}\int \frac{d \theta_j}{\pi}\sum_{m_j \in \{-1,1\}} \left(\frac{\pi}{2}\right)^23m_j\textrm{sin}^2(\theta_j)\left(\delta_{k_j,0}n_{i_j}+\delta_{i_j,0}n_{k_j}\right) \Bigg)\nonumber\\
    &\cdot \prod_{j\in \mathcal{S}''} \left(\frac{\pi^2}{8}\right)p(\{m,\vec{n}\}).
\end{align}
Here $\mathcal{S}$ is the set of qubit indices $j$ such that $i_j\neq \I$ and $k_j\neq \I$, $\mathcal{S}'$ is the set of qubit indices $j$ such that exactly one of  $i_j$ and $k_j$ is $\I$, and $\mathcal{S}''$ is the set of qubit indices $j$ such that $i_j=k_j= \I$.

In order to simplify this expression we refer back to the single-qubit results. If neither or both of $i_j$, $k_j$ are $0$, then that qubit term is proportional to the Kronecker $\delta_{i_j,k_j}$ with a coefficient from the previous section. However, for all of those qubits with only one of the two indices nonzero we have a term proportional to the expectation value. Expressing all of these conditions in terms of Kronecker $\delta$'s, we have:

\begin{align}\label{eq:delta_prime}
    \langle R'[ P_{\vec{i}}] R'[ P_{\vec{k}}]\rangle =& \prod_{j=1}^Q\Bigg( \frac{\pi^2}{8}\delta_{i_j,0}\delta_{k_j,0}+\frac{27\pi^2}{64}\delta_{i_j,1}\delta_{k_j,1} +\frac{27\pi^2}{64}\delta_{i_j,2}\delta_{k_j,2}+\frac{9\pi^2}{32}\delta_{i_j,3}\delta_{k_j,3}\Bigg)\nonumber\\
    &\;+\prod_{j=1}^Q\Bigg(\delta_{i_j,0}\Bigg(\left(\delta_{k_j,1}+\delta_{k_j,2}\right)\frac{9 \pi^2}{64}+\delta_{k_j,3}\frac{3 \pi^2}{32}\Bigg)\nonumber \\
    &\;\qquad+\delta_{k_j,0}\Bigg(\left(\delta_{i_j,1}+\delta_{i_j,2}\right)\frac{9 \pi^2}{64}+\delta_{i_j,3}\frac{3 \pi^2}{32}\Bigg)\Bigg)\Tr[\rho P_{\vec{i}}P_{\vec{k}}]\nonumber\\
    \leq& \Bigg|\prod_{j=1}^Q\Bigg( \frac{\pi^2}{8}\delta_{i_j,0}\delta_{k_j,0}+\frac{27\pi^2}{64}\delta_{i_j,1}\delta_{k_j,1} +\frac{27\pi^2}{64}\delta_{i_j,2}\delta_{k_j,2}+\frac{9\pi^2}{32}\delta_{i_j,3}\delta_{k_j,3}\nonumber\\
    &\;+\delta_{i_j,0}\Bigg(\left(\delta_{k_j,1}+\delta_{k_j,2}\right)\frac{9 \pi^2}{64}+\delta_{k_j,3}\frac{3 \pi^2}{32}\Bigg)\nonumber \\
    &\;+\delta_{k_j,0}\Bigg(\left(\delta_{i_j,1}+\delta_{i_j,2}\right)\frac{9 \pi^2}{64}+\delta_{i_j,3}\frac{3 \pi^2}{32}\Bigg)\Bigg)\Bigg|\nonumber\\
    =& \Delta'_{\vec{i},\vec{k}}
\end{align}
Here we have used the fact that $\Tr[\rho P_{\vec{i}}P_{\vec{k}}]\leq 1$ to generate this state independent bound on these products.

\subsubsection{Variance of Pole-Concentrated Estimators}

For the case of a general operator $O$, we can construct a bound of the variance based on the expectation values of products of the Pauli strings by using decomposing $O$ as
\begin{align}
    O=\sum_{\vec{i}} c_{\vec{i}} P_{\vec{i}}.  \nonumber
\end{align}
This gives us the single shot bound:
\begin{align} \label{eq:non-uniform raw bound}
    \textrm{Var}\left(\overline{R'[O]}\right)\leq\sum_{\vec{i}}\sum_{\vec{k}} |c_{\vec{i}}| |c_{\vec{k}}||\langle R'[ P_{\vec{i}}] R'[ P_{\vec{k}}]\rangle|.
\end{align}

Following~\cite{paini2021estimating}, we can then construct a state independent seminorm bound from Equation~\eqref{eq:delta_prime} by using it to bound the variance on the left hand side of Equation~\eqref{eq:non-uniform raw bound}. We then have
\begin{align}\label{eq:non-uniform bound}
    \textrm{Var}\left(\overline{R'[O]}\right)
    \leq&\sum_{\vec{i}}\sum_{\vec{k}} |c_{\vec{i}}| |c_{\vec{k}}|\Delta'_{\vec{i},\vec{k}}\nonumber\\
    =& \|O\|'^{\,2}
\end{align}
Here we define $\|O\|'=\sqrt{\sum_{\vec{i}}\sum_{\vec{k}} |c_{\vec{i}}| |c_{\vec{k}}|\Delta'_{\vec{i},\vec{k}}}$ to be the seminorm relevant to the expectation value estimation estimator $\overline{R'[O]}$.

\section{The Approximate State Formalism with Noisy Readout}

\subsection{The Effective Readout Error Channel}\label{app:effective_error_channel}
Here we consider the impact of arbitrary error channels that act immediately before or during the readout process. We will make the simplifying assumption that the single-qubit rotations immediately prior to measurement are noiseless as single-qubit gates often have higher fidelities than measurements.

As arbitrary operator expectation values can be computed as sums over Pauli strings, we will focus on how readout error channels impact the expectation value of Pauli strings.

We begin with the definition of the approximate state estimator for the Pauli string $\left\langle P_{\vec{i}} \right\rangle$:
\begin{align}\label{eq:as_in_deltas}
    \widehat{\left\langle P_{\vec{i}} \right\rangle} =& \left(\prod_{j=1}^Q \int \frac{d\vec{n}_{j}}{4\pi} \sum_{m_j \in \{-1,1\}}\right)p\left(\{m,\vec{n}\}\right)\Tr[K(\{m,\vec{n}\}) P_{\vec{i}}]\nonumber \\
    =& \left(\prod_{j=1}^Q \int \frac{d\vec{n}_{j}}{4\pi} \sum_{m_j \in \mathcal{J}_{i_j}}\right)p\left(\{m,\vec{n}\}\right)\left(\prod_{j=1}^Q m_j\left[\delta_{i_j,0}+(1-\delta_{i_j,0})3  n_{i_j}\right]\right),
\end{align}
where 
\begin{equation}
    \mathcal{J}_{i_j}= \begin{cases}
    \{-1,1\},   & \text{if } i_j > 0\\
    \{1\},          & \text{otherwise}.
\end{cases}
\end{equation}

To proceed we will perform a change of variables. Note that the rotation unitaries corresponding to measuring along the Bloch sphere directions $\{\vec{n}\}$ can have a rotation generated by $\sigma_Z$ appended at the end of the circuit without changing the measurement taken. With the addition of a normalized integration over this superfluous rotation, our integration measures become the single-qubit Haar measures. With this in mind, we then change to integrating over the Haar measure of unitaries rather than the measurement directions. With this change of variables we have
\begin{align}
    \delta_{i_j,0}+\left(1-\delta_{i_j,0}\right)n_{i} =& \frac{1}{2}Tr\left[\sigma_{i_j} U_j^\dagger\left(\sigma_z+\I\right)U_j\right]\nonumber \\
    =&\frac{1}{2}Tr\Big[\sigma_{i_j} U_j^\dagger\left(\delta_{i_j,0}\I+\left(1-\delta_{i_j,0}\right)\sigma_z\right)U_j\Big]
\end{align}
and
\begin{align}
    \left( \prod_{j=1}^Q \sum_{m_j \in \mathcal{J}_{i_j}} m_j\right)p\left(\{m,\vec{n}\}\right) =& \Tr\left[\left(\bigotimes_{j=1}^Q U_j\right)\rho\left(\bigotimes_{j=1}^Q U_j^\dagger\right)M_{\vec{i}}\right],
\end{align}
where $M_{\vec{i}}=\bigotimes_{j=1}^Q\left(\delta_{i_{j},0}\I +\left(1-\delta_{i_{j},0}\right)\sigma_z\right)$ is the observable physically measured in the absence of readout errors and $\rho$ is the state being measured.

Denoting the number of non-identity elements in $P_{\vec{i}}$ as $Q_P$, we can gather the factors of 3 from the final line of~\eqref{eq:as_in_deltas} and write
\begin{align}
    \widehat{\left\langle P_{\vec{i}} \right\rangle} =& \frac{3^{Q_P}}{2^Q}\left(\prod_{j=1}^Q \int dU_j \right)\Tr\left[ P_{\vec{i}}\left(\bigotimes_{j=1}^Q U_j^\dagger\right)M\left(\bigotimes_{j=1}^Q U_j\right)\right]\Tr\left[\rho\left(\bigotimes_{j=1}^Q U_j^\dagger\right)M\left(\bigotimes_{j=1}^Q U_j\right)\right]\nonumber\\
    =& \frac{3^{Q_P}}{2^Q}\left(\prod_{j=1}^Q \int dU_j \right)Tr\left[ P_{\vec{i}}\otimes\rho\left(\bigotimes_{j=1}^Q U_j^\dagger\right)^{\otimes 2}M_{\vec{i}}^{\otimes 2}\left(\bigotimes_{j=1}^Q U_j\right)^{\otimes 2}\right].
\end{align}
Here we have exploited the fact that a product of traces is a trace of a tensor product to re-write the expression as a single trace over two copies of the system.

Modeling readout error as an arbitrary noise channel $\mathcal{E}$ acting on the state immediately before readout (i.e. after the rotations), or equivalently as the adjoint channel $\mathcal{E}^\dagger$ acting on $M$, in the noisy case we instead have
\begin{align}
    \widehat{\left\langle P_{\vec{i}} \right\rangle} =& \frac{3^{Q_P}}{2^Q}\left(\prod_{j=1}^Q \int dU_j \right)\Tr\left[ P_{\vec{i}}\otimes\rho\left(\bigotimes_{j=1}^Q U_j^\dagger\right)^{\otimes 2}  M_{\vec{i}}\otimes \mathcal{E}^{\dagger}\left(M_{\vec{i}}\right)\left(\bigotimes_{j=1}^Q U_j\right)^{\otimes 2}\right].
\end{align}

Next we will make use of a Haar integral identity from~\cite{puchala2017symbolic}
\begin{align}\label{eq:Haar_identity}
    \int dV V_{i_1,j_1}V_{i_1',j_1'}^*V_{i_2,j_2}V_{i_2',j_2'}^*
    =&\frac{1}{d^2-1}\left(\delta_{i_1,i_1'}\delta_{i_2,i_2'}\delta_{j_1,j_1'}\delta_{j_2,j_2'}+\delta_{i_1,i_2'}\delta_{i_2,i_1'}\delta_{j_1,j_2'}\delta_{j_2,j_1'}\right)\nonumber \\
    &+\frac{1}{d(d^2-1)}\left(\delta_{i_1,i_1'}\delta_{i_2,i_2'}\delta_{j_1,j_2'}\delta_{j_2,j_1'}+\delta_{i_1,i_2'}\delta_{i_2,i_1'}\delta_{j_1,j_1'}\delta_{j_2,j_2'}\right).
\end{align}
In order to apply this identity we first expand the integral over $U_1$ as
\begin{align}
    \widehat{\left\langle P_{\vec{i}} \right\rangle} = \frac{3^{Q_P}}{2^Q}\left(\prod_{j=2}^Q \int dU_j \right) \int dU_1& U_{1_{i_1,j_1}}U_{1_{i_1',j_1'}}^*U_{1_{i_2,j_2}}U_{1_{i_2',j_2'}}^* \nonumber\\
    & \Tr\Bigg[P_{\vec{i}}\otimes\rho\left(|i_1\rangle\langle j_1|\otimes\bigotimes_{j=2}^QU_j^\dagger\otimes |j_1'\rangle\langle i_1'|\otimes\bigotimes_{j=2}^Q U_j^\dagger\right)\nonumber \\
    &\qquad M_{\vec{i}}\otimes \mathcal{E}^{\dagger}\left(M_{\vec{i}}\right)\left(|i_2\rangle\langle j_2|\otimes\bigotimes_{j=2}^QU_j\otimes |j_2'\rangle\langle i_2'|\otimes\bigotimes_{j=2}^Q U_j\right)\Bigg]. 
\end{align}
Using~\eqref{eq:Haar_identity} to perform the $U_1$ integral results in
\begin{align}
    \widehat{\left\langle P_{\vec{i}} \right\rangle} = \frac{3^{Q_P}}{2^Q}\frac{1}{3-2\delta_{i_j,0}}\left(\prod_{j=2}^Q \int dU_j \right)\Tr\Bigg[&\Tr_{1,1'}\left[P_{\vec{i}}\otimes\rho \mathcal{P}_{1,1'}\right]\left(\bigotimes_{j=2}^QU_j^\dagger\otimes \bigotimes_{j=2}^Q U_j^\dagger\right)\nonumber\\
    &\Tr_{1,1'}\left[M_{\vec{i}}\otimes \mathcal{E}^{\dagger}\left(M_{\vec{i}}\right)\mathcal{P}_{1,1'}\right]\left(\bigotimes_{j=2}^QU_j\otimes \bigotimes_{j=2}^Q U_j\right)\Bigg],
\end{align}
where $\mathcal{P}_{j,\ell}$ is the swap operator for qubits $j$ and $\ell$. Repeated application of this identity for the remaining qubits results in
\begin{align}\label{eq:error_suppression}
    \widehat{\left\langle P_{\vec{i}} \right\rangle} =& \frac{1}{2^Q}\Tr\left[M_{\vec{i}} \otimes \mathcal{E}^{\dagger}\left(M_{\vec{i}}\right)\left(\prod_{j=1}^Q\mathcal{P}_{j,j'}\right)\right]\Tr\left[P_{\vec{i}}\otimes\rho\left(\prod_{j=1}^Q\mathcal{P}_{j,j'}\right)\right]\nonumber\\
    =& \frac{1}{2^Q}\Tr\left[M_{\vec{i}}  \mathcal{E}\left(M_{\vec{i}}\right)\right]\Tr\left[P_{\vec{i}}\rho\right].
\end{align}
Thus, we see that the effect of readout error is to suppress expectation values by a factor of $\frac{1}{2^Q}\Tr\left[M_{\vec{i}}  \mathcal{E}\left(M_{\vec{i}}\right)\right]$. 

Note that this suppression is symmetric with respect to changes in $P_{\vec{i}}$ that result in the same set of qubits being acted on with identity operators. Given that degree of symmetry, the effective error channel can be faithfully described as a composition of depolarizing channels acting on different subsets of the qubits.

\subsection{Relationship to Twirling}\label{app:twirl}
It has been claimed, without proof, in previous works that randomised readout is a twirling operation~\cite{karalekas2020quantum,chen2021robust}. A twirl $\tau_G(\cdot)$ is a super-super-operator, that is an operator that acts on channels. The nature of the twirl is dependent on the selection of a unitary group representation $G$, which we say that we twirl with. When G is discrete we have
\begin{equation}
    \tau_G\left(\mathcal{E}\left(\cdot\right)\right)= \frac{1}{|G|}\sum_{U\in G} U^\dagger\mathcal{E}\left(U \cdot U^\dagger\right)U,
\end{equation}
where $|G|$ is the cardinality of that $G$. If $G$ is continuous this is instead
\begin{equation}
    \tau_G\left(\mathcal{E}\left(\cdot\right)\right)= \int_G dU \; U^\dagger\mathcal{E}\left(U \cdot U^\dagger\right)U.
\end{equation}

In the following we will make use of the Pauli Transfer Matrix (PTM) representation of channels. Denoting the normalized vectorization of the Pauli operator $\sigma_j$ as $|\sigma_j\rangle\rangle$, we consider the components of the PTM representation of $\mathcal{E}$ as:
\begin{align}
    [\mathcal{E}]_{\vec{i},\vec{j}} =& \langle\langle P_{\vec{i}} | \mathcal{E} |P_{\vec{j}} \rangle\rangle\nonumber \\
    =& \frac{1}{2^Q}\Tr[P_{\vec{i}} \mathcal{E}(P_{\vec{j}})].
\end{align}

We now specialize to the case where $G$ is a tensor product of single-qubit $t\geq2$-designs as that case commonly used in the classical shadows formalism. (We note that the tetrahedral group used in the main text is the minimal group that meets this criterion.) As acting on a Pauli string $P_{\vec{i}}$ with a tensor product of single-qubit unitaries leaves all single-qubit identity operators unchanged, the actions of elements of such a tensor product $G$ is closed on the subspace spanned by the set of basis vectors $|P_{\vec{i}}\rangle\rangle$ that have identity operators in the same place, denoted as $\mathcal{S}_{\vec{i}}$. Further, if $G$ is a tensor product of $t\geq2$-designs, such as the tetrahedral or Clifford groups, there are elements of $G$ that can map any $P_{\vec{i}}$ to any other $P_{\vec{j}}\in \mathcal{S}_{\vec{i}}$. We therefore have that if $G$ is a tensor product of $t\geq2$-designs, then $G$ is an irreducible representation on $\mathcal{S}_{\vec{i}}$.

If $G$ is irreducible on $\mathcal{S}_{\vec{i}}$, then we have that for any $P_{\vec{i}},P_{\vec{j}}\in\mathcal{S}_{\vec{i}}$
\begin{align}\label{eq:twirled_suppression}
    \langle\langle P_{\vec{i}}|\tau_G\left(\mathcal{E}\right)|P_{\vec{j}}\rangle\rangle=& \frac{1}{2^Q |G|}\Tr\left[\sum_{U\in G} P_{\vec{i}} U^\dagger\mathcal{E}\left(U P_{\vec{j}} U^\dagger\right)U\right]\nonumber \\
    =& \frac{1}{2^Q}\left(\frac{1}{\textrm{dim}(\mathcal{S}_{\vec{i}})}\sum_{\big\{\vec{k} \big||P_{\vec{k}}\rangle\rangle \in \mathcal{S}_{\vec{i}}\big\} }\Tr[P_{\vec{k}}\mathcal{E}\left(P_{\vec{k}}\right)]\right)\Tr[P_{\vec{i}}P_{\vec{j}}].
\end{align}
The second line follows from applying Assertion 1 of ~\cite{paini2000quantum} to $\mathcal{E}$ in the space of super-operators on $\mathcal{S}_{\vec{i}}$. Here $dim(\mathcal{S}_{\vec{i}})=3^{Q_P}$ is the dimension of $\mathcal{S}_{\vec{i}}$ when the number of non-Pauli elements in $P_{\vec{i}}$ is $Q_P$. As an aside, we note that it is possible to derive the structure of the effective channel of the approximate state (i.e. Equation~\eqref{eq:error_suppression}) using the same irreducibly arguments used to derive Assertion 1 of ~\cite{paini2000quantum}. This approach is only valid for groups and their representations that guarantee the irreducibility in $\mathcal{S}_{\vec{i}}$, which is the case for the tetrahedral group and its supergroup (equivalent to the conditions of 2-designs).

We now compare this result with the one for the approximate state formalism. As the approximate state formalism's effective channel only depends on a single element diagonal element of the matrix $[\mathcal{E}]_{\vec{i},\vec{j}}$  (the one corresponding to $|M_{\vec{i}}\rangle\rangle$) rather than an average over the diagonal elements corresponding to all of the Pauli string basis vectors in $\mathcal{S}_{\vec{i}}$, the right hand sides of  Equations~\eqref{eq:error_suppression} and~\eqref{eq:twirled_suppression} can be quite different. Therefore, despite the previously published assertions,  while random readout is very similar to twirling the error channel $\mathcal{E}$ the coefficients are different.
\end{document}